\documentclass{aa}  

\usepackage{graphicx}
\usepackage{longtable}
\usepackage{epsfig,amsmath,wrapfig}
\usepackage{lscape}
\usepackage{txfonts}
\usepackage{xcolor}
\usepackage{ulem}

\newcommand{\dF}{{^{^*}\!\!F}}
\newcommand{\bP}{{\bf P}}
\newcommand{\bF}{{\bf F}}
\newcommand{\bU}{{\bf U}}
\newcommand{\bS}{{\bf S}}

\newcommand{\sB}{{\mathcal{B}}}
\newcommand{\sR}{{\mathcal{R}}}
\newcommand{\sQ}{{\mathcal{Q}}}

\newcommand{\del}{{\partial}}
\newcommand{\tu}{{\tilde{u}}}

\usepackage{natbib}

\begin{document} 

\title{Chemical evolution and kilonova implications of post-merger accretion disk winds}

\titlerunning{Heavy-element nucleosynthesis in GRB engines}

   \author{Agnieszka Janiuk
          \inst{1}\fnmsep\thanks{agnes@cft.edu.pl}
          \and
          Joseph Saji \inst{1}
           \and
          Gerardo Urrutia\inst{1}
          }

              \institute{Center for Theoretical Physics, Polish Academy of Sciences, Al. Lotnik\'ow 32/46, 02-668 Warsaw, Poland\\
              }

  \abstract
    {Several gamma-ray bursts (GRBs) have recently been associated with kilonova emission. We study the mechanisms that account for this effect by means of the radioactive decay of elements synthesized in accretion disk winds.}
   {We modeled r-process nucleosynthesis in the accretion disk wind system associated with the prompt GRB phase. 
   We computed the time-dependent general relativistic  magnetohydrodynamical (GRMHD) evolution of a GRB central engine in which the newly formed black hole accretes mass from the post-merger remnant. 
   We explored the wind properties over a range of initial system parameters and study representative cases of compact binary merger progenitors. }
    {We computed a suite of 2D and 3D GRMHD numerical simulations with a tabulated three-parameter equation of state that allows for the evolution of chemical composition in the accretion flow.
   We accounted for neutrino emission by incorporating a leakage scheme in which the neutrino optical depth was calculated along the radial rays. We parametrized the optically thick and thin tori with different values of the pressure maximum and disk entropy, while we parametrized the strength of large-scale poloidal magnetic fields according to the chosen gas-to-magnetic pressure ratio. To probe the winds, we followed particle trajectories.

From these simulations, we derived the nucleosynthetic yields of heavy elements in the outflows, and we mapped the regions of lanthanide-rich and lanthanide-poor ejecta. 
} 
   {We find that the outflow carries a high-mass of neutron-rich material expanding at mildly relativistic velocities. Our accretion disks operating in the standard and normal accretion (SANE) mode can power GRB jets via neutrino annihilation if the disk-to-black hole mass ratio exceeds approximately 0.01 and the black hole spins rapidly. Slowly spinning black holes surrounded by massive post-merger disks can also power these jets and can serve as sites of efficient lanthanide nucleosynthesis, which is responsible for the red or purple kilonova components. Long-duration GRBs are potentially produced by a specific type of merger event with an eccentric orbit, but in this case, their lanthanide fraction is reduced, and the GRB luminosity due to neutrinos is also weak.
   We find that none of the weakly magnetized post-merger models considered here can reproduce the bright kilonovae associated with long GRBs such as 211211A}
   {}

   \maketitle
%

\section{Introduction}

Gamma-ray bursts (GRBs) are sudden releases of approximately $10^{51}-10^{54}$ erg of energy within a radius of less than 100~km, lasting from 0.01 to 100~s (for reviews, see e.g., \cite{Piran2004}; \cite{Kumar2015}). Based on the duration $T_{90}$, which is defined as the time interval over which $90\%$ of the total background-subtracted counts are observed, GRBs are usually separated into two classes: long GRBs (LGRBs; $T_{90}>2 s$), which originate from the core collapse of massive stars \citep{Woosley1993,Hjorth2003}, and short GRBs (SGRBs; $T_{90}<2 s$), which originate from the coalescence of binary neutron stars (BNS) or neutron star-black hole (BHNS) binary systems \citep{Eichler1989, Narayan1992}.

Most observed GRBs ($70 \%$) have durations greater than two seconds and are classified as LGRBs. Because these events constitute the majority of the population and tend to have the brightest afterglows,
they have been studied in greater detail than their short counterparts.
Almost every well-studied LGRB has been linked to a galaxy undergoing rapid star formation, and in many cases to a core-collapse supernovae (CCSN) as well \citep{Woosley2006}. Observations of LGRB afterglows, which reflect their association with high redshift ($\gtrsim 5$), are also consistent with their origin in star-forming regions \citep{Pontzen2010}. 

The leading hypothesis for the origin of short GRBs is that they result from mergers of compact binary systems, such as two neutron stars. This theory is supported by the detection of gravitational waves from a neutron star merger (GW170817) coinciding with an SGRB \citep{2017ApJ...848L..13A}.
The environments of SGRBs are diverse. Some short GRBs originate in older, elliptical galaxies with little or no ongoing star formation, indicating that their progenitors can have long lifetimes before merging 
\citep{2011NewAR..55....1B}.
However, the majority of these GRBs are found in star-forming galaxies, which differ from the hosts of long-duration events. These galaxies typically have lower star formation rates and higher metallicities, resembling the general field galaxy population.
In addition, short GRBs occur often at significant offsets from the centers of their hosts, sometimes even in the galactic halos. This distribution aligns with predictions of neutron star merger rates, as the neutron stars can receive ``kicks'' during their formation, propelling them far away from their birthplaces (see e.g., \cite{2018MNRAS.481.4009V, 2018MNRAS.480.2011G}).
A wide range of progenitor ages, indicated by the diversity of spatial distributions, suggests that the time between the formation of a progenitor system and its merger can vary significantly, from millions to billions of years \citep{2022ApJ...940...56F}.

Compact binary mergers can eject large amounts of neutron-rich material, which serve as sites of r-process nucleosynthesis and the formation of heavy elements \citep{1974ApJ...192L.145L}.
The ejecta then produce radiation emitted by the radioactive decay of the unstable heavy isotopes, which powers an optical-infrared transient called a kilonova. As predicted by \cite{1998ApJ...507L..59L}, a tremendous amount of energy, together with heavy elements, is emitted in an explosive outflow. Elements such as gold and platinum are produced in r-process nucleosynthesis, which requires extremely high neutron densities.

The elemental composition of Earth and the Solar System is now attracting increasing interest in fundamental questions of nuclear astrophysics\citep{2023A&ARv..31....1A}.
The Solar System's abundance pattern results from multiple generations of stellar nucleosynthesis.
A considerable amount of $^{60}$Fe (half-life $\sim$2.6 Myr) 
from deep-ocean and lunar samples
likely originated from multiple supernova explosions within $\sim$100 parsecs during the last 10 Myr. 
Supernova deposition is also primarily responsible for synthesizing elements slightly beyond iron, such as $^{56}$Ni, which later decays to  $^{56}$Fe, one of the dominant elements in planetary cores of the Solar system.
In addition, traces of the heaviest elements, including lanthanides and actinides such as thorium, uranium, and plutonium, are found on Earth. These elements are produced primarily in compact binary mergers and, less frequently, in rare magneto-rotational \citep{Nishimura_2015,10.1093/mnras/stac3185} or collapsar-type supernovae. The presence of a small amount of $^{244}$Pu on Earth suggests that a kilonova event may have occurred within the last $\sim$100 Myr in the vicinity of the Solar System \citep{Wang_2021}.

One of the first suspected kilonova events was the gamma-ray burst GRB 130603B, detected by the Swift and Hubble Space telescopes. It was a short event, lasting less than 2~s , with a faint red afterglow observed several days after the initial burst. However, the emission was much brighter and more persistent than expected for a typical GRB afterglow \citep{2013Natur.500..547T}. Its red color and light curve slope were attributed to the radioactive decay of newly synthesized heavy elements 
\citep{2013ApJ...774L..23B}. The observation thus indicated a binary neutron star merger environment.
The first confirmed GRB associated with a kilonova was observed in 2017 \citep{2017ApJ...848L..17C}. The kilonova light faded over a couple of weeks, making it shorter-lived than any supernova. The spectrum evolved from blue (hot) to red (cooler) \citep{2017ApJ...848L..12A}.

The source detected in 2021 was a kilonova associated with LGRB 211211A. It is interpreted as a merger between a neutron star (NS) of approximately 1.23 solar masses and a black hole (BH) of 8.21 solar masses, with an aligned spin of 0.62. The data appear to support an NS-first-born NS-BH formation channel \citep{Zhu_2022};
 however, we notice that the gravitational-wave (GW) parameters are degenerate and do not allow a unique identification of the pre-merger system.
In this case, the potential re-brightening of the emission may be influenced by late-time fallback accretion, while the ejected material is significantly more massive than in other progenitor types (at least $10^{-3} M_{\odot}$).
As an alternative to this scenario, some authors propose a neutron star-white dwarf (NS-WD) merger  \citep{2022Natur.612..232Y, Zhong_2023}, in which the toroidal magnetic field of the NS is amplified by accretion of material from the white dwarf. This process leads to prompt GRB emission, while the extended phase occurs during the propeller regime \citep{2022ApJ...934L..12G}. In this case, the kilonova is powered by the decay of Ni-56 in the debris ejected from the WD.

Furthermore,
GRB 230307A is the second-brightest long-duration GRB ever detected, and exhibits a rare association with a kilonova. Theoretical interpretations of this event range from a compact binary merger \citep{Levan2024, Liu_2025} to alternative scenarios involving a collapsar \citep{2025arXiv250903003R} or an NS-
WD merger with a long-lived magnetar engine \citep{Wang_2024, Arunachalam_2025}. 
The possibility of a kilonova originating from a collapsing massive star, with angular momentum slightly lower than that of a typical LGRB progenitor, has also been discussed \citep{2019Natur.569..241S}. The r-process-enriched collapsar model has been used to fit the light curves of GRB 2011211A; however, the fits they exhibit high parameter degeneracies and require exceptionally high ejecta velocities \citep{Barnes_2023}.

In this work, we explore the scenario of a kilonova driven by NS-NS and NS-BH mergers. In particular, we investigate the range of energetics and durations of the events, as well as the properties of dynamical ejecta launched from the post-merger accretion disk. We also investigate the possible formation of delayed accretion episodes, which may play a key role in sustaining the jet of a long-duration GRB launched after the compact binary merger.

Our methodology is based on detailed numerical simulations using new version of the HARM-EOS code.
The code (new branch) was developed based on our publicly available HARM-COOL\footnote{https://github.com/agnieszkajaniuk/HARM\_COOL} scheme and was updated by adding a  three-parameter equation of state (EOS) and a neutrino leakage module.
These updates aim to correctly represent the influence of chemical evolution on the engine properties and to account for neutrino viscosity.
We introduced a poloidal magnetic field component in the accretion flow, and studied the influence of its strength on the magneto-rotational instability. We performed both 2D and 3D numerical simulations for a period long enough to launch a sufficient amount of dynamical ejecta.

The article is organized as follows. In Section \ref{sect:model}, we describe the model and the setup of our numerical simulations. 
In Section \ref{sect:results}, we present our calculations, describing the initial conditions, time evolution of BH torus systems, 
  and the results of r-process nucleosynthesis calculations. 
  We discuss our results in Section \ref{sect:discussion}and provide a summary and conclusions in Sect. \ref{sec:conclusions}.

\section{Numerical model}
\label{sect:model}

\subsection{Basic equations}

We used a developed version of a general relativistic magnetohydrodynamic code, based on the high-accuracy relativistic magnetohydrodynamics (HARM) framework, which evolved from the code made publicly available by \cite{Gammie2003} and described in detail by \cite{Noble2006}. The code implements a conservative, shock-capturing scheme with low numerical viscosity to solve the hyperbolic system of general relativistic  magnetohydrodynamical (GRMHD) partial differential equations. The numerical scheme uses the plasma energy-momentum tensor, $T^{\mu \nu}$, with contributions from matter (gas) and the electromagnetic field. For the GRMHD  evolution, we solved two fundamental equations: the equation of mass and
angular momentum conservation, and the energy equation,

\begin{equation}
T_{(m)}^{\mu \nu}=\rho hu^{\mu}u^{\nu}+pg^{\mu \nu}
\end{equation}

\begin{equation}
T_{(em)}^{\mu \nu}=b^{k}b_{k} hu^{\mu}u^{\nu}+\frac{1}{2} b^{k}b_{k}g^{\mu \nu}-b^{\mu}b^{\nu}
\end{equation}

\begin{equation}
T^{\mu \nu}=T_{(m)}^{\mu \nu}+T_{(em)}^{\mu \nu},
\end{equation}
where $u^{\mu}$ denotes the four-velocity of gas, 
$b^{\mu}$ is the magnetic four-vector, and $h$ is the fluid specific enthalpy. The continuity and energy-momentum conservation equations are written as 
\begin{equation}
(\rho u^{\mu})_{;\mu}=0
\label{eq:cont}
\end{equation}
\begin{equation}
T_{\nu;\mu}^{\mu}=\sQ u_{\nu}
\label{eq:ener},
\end{equation}
where $\sQ$ is the energy change per unit volume due to neutrino cooling.
Finally, the Maxwell equations in the limit of ideal MHD are
\begin{equation}
^{*}F^{\mu \nu}_{;\mu}=0,
\end{equation}
where $F^{\mu \nu}$ denotes the electromagnetic (Faraday) tensor.

We wrote the GRMHD equations in conservative form by implementing a Harten-Lax-van Leer (HLL) solver \footnote{This is an approximate Riemann solver. In general, Riemann solvers are used to solve hyperbolic partial differential equations based on the solutions of the corresponding Riemann problem. They compute the numerical flux across a discontinuity for a conservation equation with piecewise-constant initial data that has a single discontinuity in the domain of interest). They form an important part of high-resolution schemes in computational fluid dynamics and computational magnetohydrodynamics.}\citep{Harten1983} to calculate the corresponding fluxes numerically .

For this study, we developed a new version of our code, based on that introduced by \cite{Janiuk2019}.
The new version, now named HARM\_EOS,
includes a scheme for simulating the evolution of a post-merger accretion disk with detailed microphysics from equation of state (EOS). It uses a tabulated three-parameter equation of state for dense matter and accounts for neutrino cooling and heating via a leakage scheme.

In the continuity equation \ref{eq:cont}, the conserved quantity is the baryon density, $\rho=m_{b}n_{b}$. The lepton density changes due to weak interactions and is determined by the net rate of neutrino and antineutrino number density, $\sR$, as
\begin{equation}
  (\rho Y_{e} u^{\mu});_{\mu} = \sR.
\end{equation}
Here, $Y_{e} = n_{e}/n_{b} = m_{b}n_{e}/\rho$ denotes the electron fraction.

With this additional constraint, the numerical scheme must solve the GRMHD equations in conservative form, which requires rebuilding the inversion scheme to recover the conserved variables.
The recovery methods tested and implemented in our code, as well as the neutrino treatment details, are presented in Appendices \ref{sect:recovery} and \ref{sect:neutrino}.

To investigate accretion disk outflows, we used the tracer particle method, as implemented in \cite{Janiuk2019}. The particles are initialized in the densest regions of the initial accretion disk, and record information about the outflow physical properties over time.

In the following, we discuss the initial conditions of the models evolved by our numerical scheme and present the recovery method included in our code. We then show the results of numerical simulations of  post-merger systems following the coalescence of either binary neutron stars or a neutron star with a stellar-mass BH.
We find that the plasma is highly neutron-rich, and the r-process nucleosynthesis in the ejected material may lead to the creation of unstable heavy isotopes responsible for powering a kilonova.

We performed our simulations in an axisymmetric 2D setup, and we also ran selected models in 3D. We adopted resolutions of $256 \times 256 \times 1$ and $256 \times 256 \times 64$, in radial, polar, and azimuthal directions. We analyzed the ejecta from the post-merger accretion disk using test particles (tracers) to determine the mass and composition of the kilonova-powering material.
We then studied nucleosynthesis of heavy elements in the post-merger winds via postprocessing of the tracer results.
In one of the 3D runs, we also introduced a non-axisymmetric perturbation in the initial  internal energy distribution within the torus, imposed at 0.4\%.

\subsection{Initialization of the run}

We began our simulations from a pressure-equilibrium configuration of a rotating torus around a Kerr BH.
We solved for its structure given the specific enthalpy, $h$, resulting from the Fishbone-Moncrief (FM) torus angular-momentum solution $l$ \cite{Fishbone76}. 
We obtained the distribution of density and temperature in the disk using the Newton-Raphson method, which minimizes the condition
$h=1+P/\rho + u/\rho$, where $P$ and $u$ denote the pressure and internal energy obtained from the Helmholtz EOS tables
\citep{2000ApJS..126..501T}.
The fixed parameters are the entropy and electron fraction in the disk,
$S_{disk}$ and $Y_{e, disk}$.

The initial torus extends from the cusp radius, $r_{in}$, to the radius of maximum pressure, $r_{max}$, as defined by the FM solution. Outside the torus, we set constant density and temperature conditions, as well as a constant electron fraction.

We seeded the torus with an initial poloidal magnetic field. We assumed the form of the magnetic vector potential 
given by
\begin{equation}
A_{\varphi}=\max \left({\bar{\rho} \over \rho_{\rm max}} - \rho_{0},\, 0\right),
\end{equation}
where $\bar{\rho}$ is the density in the torus (averaged over two neighboring cells) and 
$\rho_{\rm max}$ is the density maximum. We adopted $\rho_{0}=0.2$ to restrict the initial field to regions where the torus density exceeds $0.2 \rho_{\max}$.
Other spatial components of the initial vector potential were $A_{r}=A_{\theta}=0$; therefore, the magnetic field has only $B_{r}$ and $B_{\theta}$ components.
We set the field strength by normalizing it to the chosen minimum gas-to-magnetic pressure ratio, 
$\beta_{min}$, within the torus. Our typical values $\beta_{min}$ were in the range $50-100$ to ensure that the initial torus was weakly magnetized.

The torus was embedded in a cool, rarefied atmosphere. We assumed an atmospheric density of $D_{atm} = 3 \times 10^{-7}$ g cm$^{-3}$, which exceeds the numerical density floor (typically set at $10^{-5}$ in code units and decreasing with radius as $r^{-2}$), 
and we adopted an atmospheric electron fraction of $Y_{e, atm}=0.45$. The atmospheric temperature was then obtained from the EOS tables and is approximately $T_{atm}=0.002$ MeV.
The floor temperature was much lower and scales with radius as $T_{floor}=T_{0}/r$, with $T_{0}=10^{3}$ K. 

\section{Results}
\label{sect:results}

We explored the properties of two fiducial tori with masses of $\sim 0.1 M_\odot$ and  $\sim 0.01 M_\odot$, optically thick or thin to neutrinos, respectively, and located around a $3 M_\odot$ 
BH mass formed after a compact binary merger. These models represent typical outcomes of a binary NS-NS merger scenario. 
 We also investigated the scenario of a higher mass ratio between the BH and disk mass, considering lower remnant tori of   $\sim 0.02 M_\odot$ and  $\sim 0.002 M_\odot$, for the optically thick and thin cases, respectively.
 These models aimed to reconstruct BH-NS merger scenarios and quantify differences in the nucleosynthesis patterns between the two merger types.
Our choice of setups are validated for typical cases in which merger occurs at high mass-ratio and/or with a slowly spinning BH. Low mass-ratio high-spin BH–NS systems can also produce massive disks. 
 
 Finally, we studied a system of a $\sim 0.8 M_{\odot}$ post-merger torus located around an 8.2 $M_{\odot}$ BH, motivated by recently observed kilonovae and constraints on their progenitor parameters.
 Therefore, while we assumed a rapidly spinning BH ($a \sim 0.9$) for fiducial models, we adopted a moderate BH spin value ($a \sim 0.6$) for the massive torus case.

The fiducial mode had a BH mass of $3 M_{\odot}$ and a dimensionless spin $a=0.9375$. The initial gas-to-magnetic pressure ratio, 
$\beta$, was normalized to the minimum value within the pressure maximum radius, $r_{max}$.
The inner disk radius was initially located at $r_{in}=4 r_{g}$.
The disk mass ranged from  $0.01 - 0.1 M_{\odot}$ for the optically thick and thin models, respectively.
The optical thickness of the torus was regulated by the choice of entropy per baryon, $S_{disk}$, as the values of density and temperature in the torus were determined from EOS tables for a given entropy, electron fraction, and specific enthalpy profile. To achieve this, we prescribed the initial chemical composition of the torus with a constant electron fraction, $Y_{e}=0.1$. The entropy per baryon was $S=10$ for the optically thick case and $S=7$ for the optically thin case.

In addition to the fiducial models, we also ran 
test models with lower disk masses of $0.02-0.002 M_{\odot}$. 
Finally, we simulated
the system with parameters tuned to the central engine of the kilonova event
GRB 211211A,
for which the BH spin and mass-ratio estimates were obtained from observations. To test whether this kilonova scenario could represent a long duration GRB event, we modified the initial condition to account for a potentially eccentric merger by introducing a non-axisymmetric perturbation in the energy distribution.

Circularization of material from a tidally disrupted companion occurs through self-intersecting shocks, accompanied by relativistic apsidal precession, with thermal energy damped at specific azimuths, rather than uniformly 
\citep{2016MNRAS.455.2253B}.
However, non-axisymmetric modes can also be produced within the disk via generic instabilities, such as Rossby waves or Papaloizou-Pringle instabilities 
\citep{2011A&A...532A..30K}. Nevertheless, in compact binary mergers, strong azimuthal structure may result from tidally disrupted companions, even when the orbital motion is quasi-circular. 
If the binary is genuinely eccentric, which can occur, for example, if it formed in a dense star cluster by a capture mechanism or in a triple system, then episodic mass transfer will occur at subsequent pericenter passages, prior to the star's disruption, leaving imprints on the disk structure. This scenario is more likely for BH-NS mergers with large mass ratios \citep{2025ApJ...978..126Z}.
To verify the original eccentricity of the system independently of disk morphology, additional information must be considered, such as the burst-like shape of the GW signal \citep{2012PhRvD..85l4009E}. The parameters and final times of all runs are given in Table \ref{tab:Sim_Models}. 

\begin{table*}[h]
 \caption{Simulation models and parameters}
  \centering
  \begin{tabular}{cccccccccccc}
    \hline
    Model & resolution & $r_{in}$ & $r_{max}$ & $\beta$ &  $a$ & $M_{BH}$ & $M_{disk}$ & $S_{disk}$ & $N_{tr} (t_{0})$ & $N_{tr}^{out} (t_{f})$  & $ t_{f}$ \\ \hline
    \hline
    2D-Thick-s & 288x256x1 & 4.0 & 9.0 & 100 & 0.9375 & 3.0 & 0.021 & 7 & 11072 & 186 & 20,000\\ 
    2D-Thin-s & 288x256x1 & 4.0 & 9.0 & 50 & 0.9375 & 3.0 & 0.002 & 10 & 11049 & 372 & 20,000\\ 
    2D-Thick & 288x256x1 & 4.0 & 9.78 & 100 & 0.9375 & 3.0 & 0.129 & 7 & 15998 & 1037 & 20,000\\ 
    2D-Thin & 288x256x1 & 4.0 & 9.78 & 50 & 0.9375 & 3.0 & 0.013 & 10 & 15988 & 554 & 18,200\\ 
    2D-kn & 288x256x1 & 4.0 & 12.6   &   50  &  0.6 & 8.2 & 0.785 & 10 & 20392 & 931 & 20,000\\
    \hline 
    3D-Thick & 256x128x64 & 4.0 & 9.78 & 100 & 0.9375 & 3.0 & 0.129 & 7 & 8131 & 1063 & 19,800\\ 
     3D-Thin & 256x128x64 & 4.0 & 9.78 & 50 & 0.9375 &  3.0 & 0.013 & 10 & 8124 & 544 & 18,300 \\ 
     3D-kn & {288x128x64} & 4.0 & 12.6   &   50  &  0.6 & 8.2 & 0.781 & 10 & 10344 & 1000 & 20,000\\ 
    3D-kn-p & {288x128x64} & 4.0 & 12.6   &   50  &  0.6 & 8.2 & 0.834 & 10 & 9954 & 281 & 20,000 \\ 
    \hline
    \hline
    \\
  \end{tabular}
   \label{tab:Sim_Models}
\end{table*}

\subsection{Time-dependent evolution}
\label{sect:time_evol}

\subsubsection{Fiducial thin and thick models}\label{sec:fiducialThinAndThick}

\begin{figure}
    \includegraphics[width=0.49\textwidth]{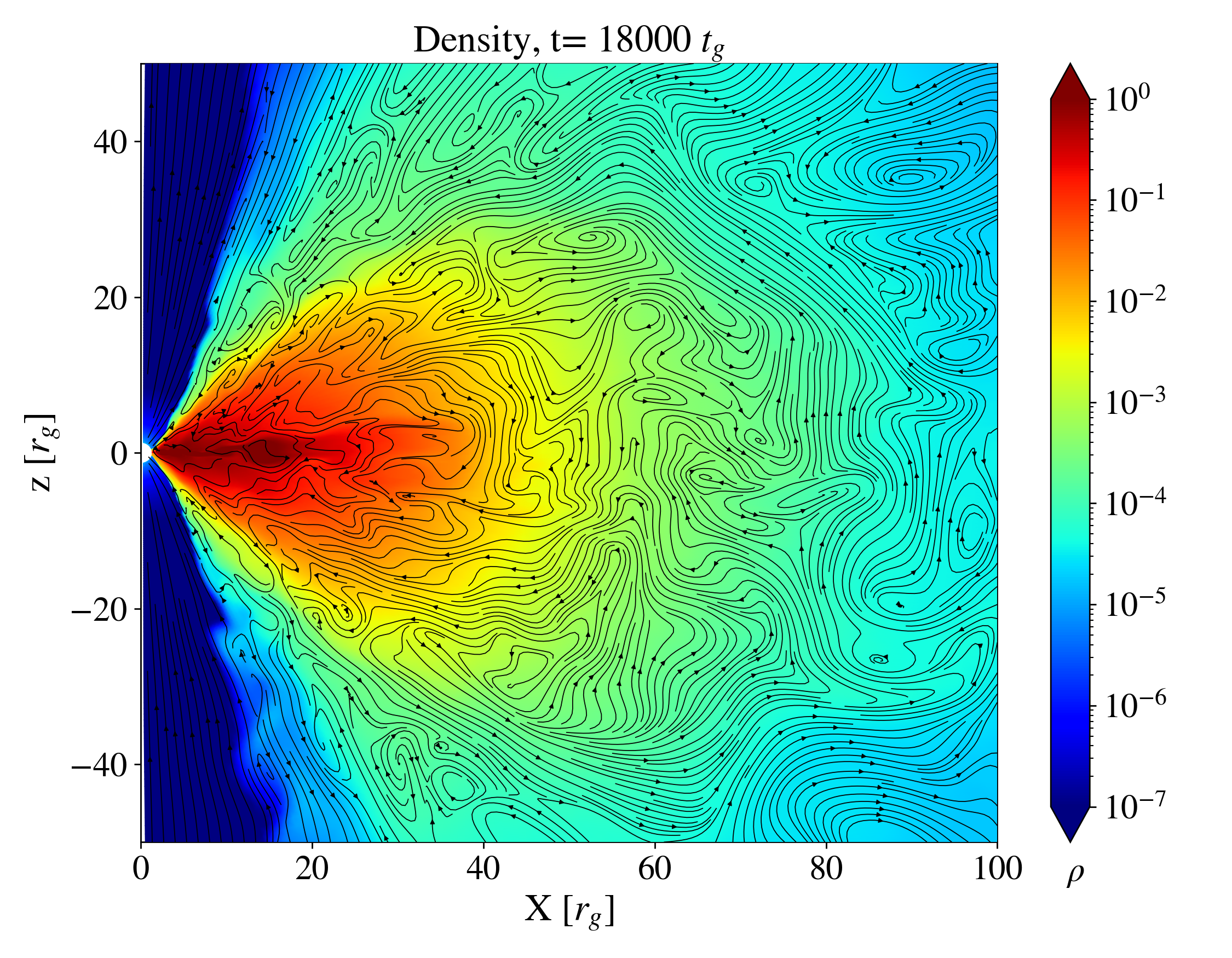}
    \includegraphics[width=0.49\textwidth]{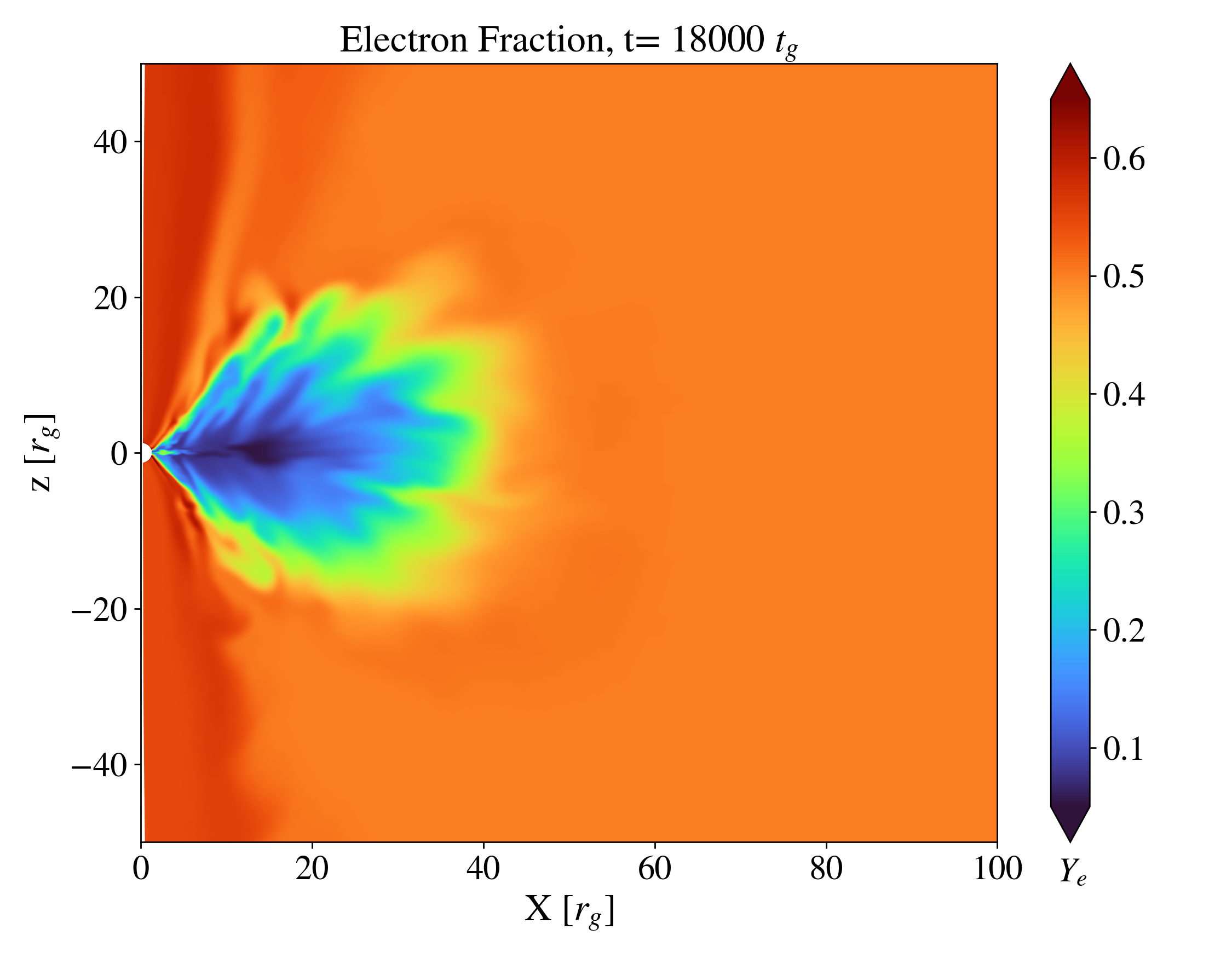}
    \caption{Snapshots from the 2D simulation of the GRB central engine for model 2D-Thin.
      The top panel shows the density distribution and magnetic-field 
    streamlines in the innermost part of the accretion disk, while the bottom panel shows the electron fraction. Snapshots are taken at $t=18000 ~t_{\rm g}$ }
    \label{fig:GRrun}
  \end{figure}

\begin{figure}
    \includegraphics[width=0.49\textwidth]{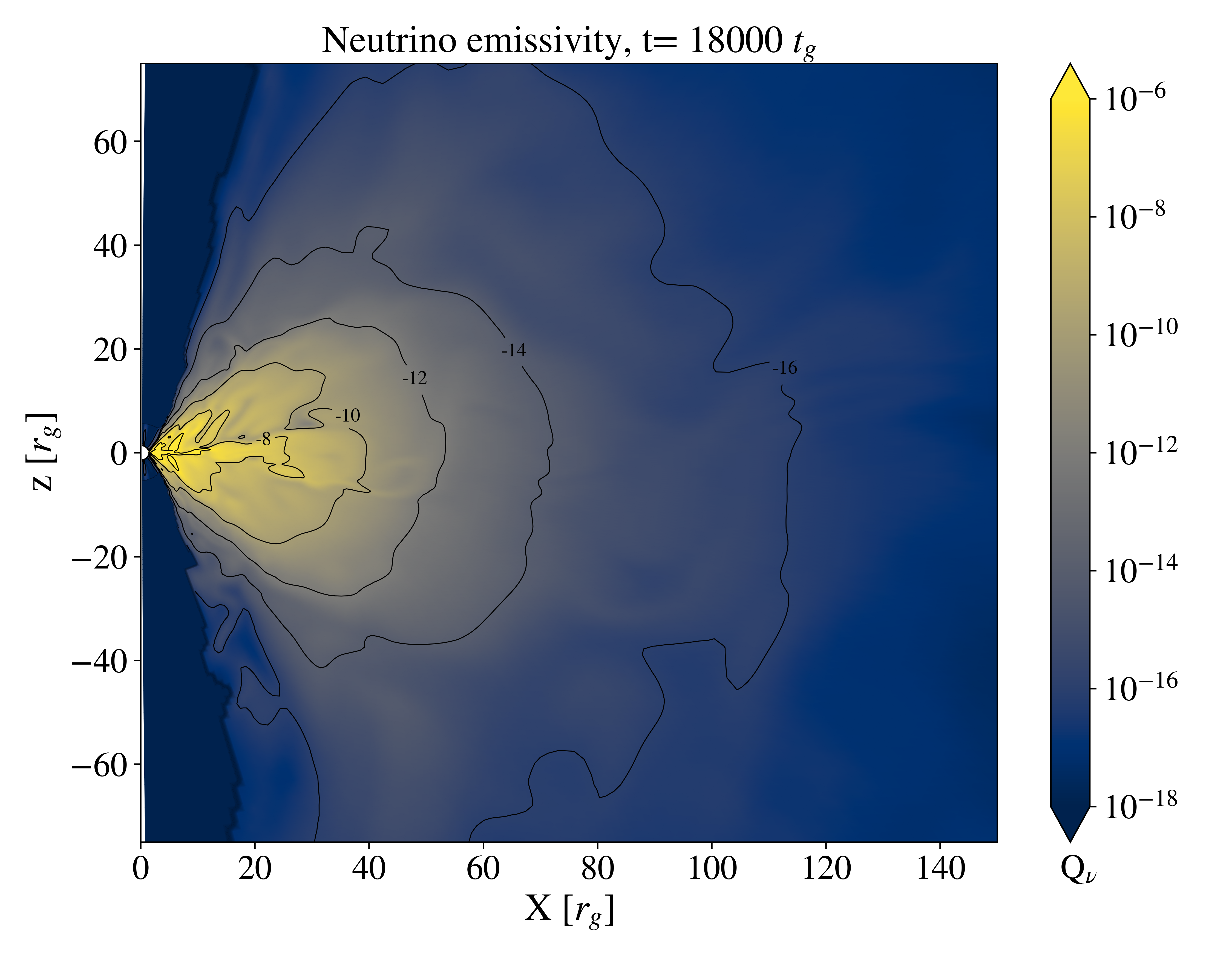}
    \includegraphics[width=0.49\textwidth]{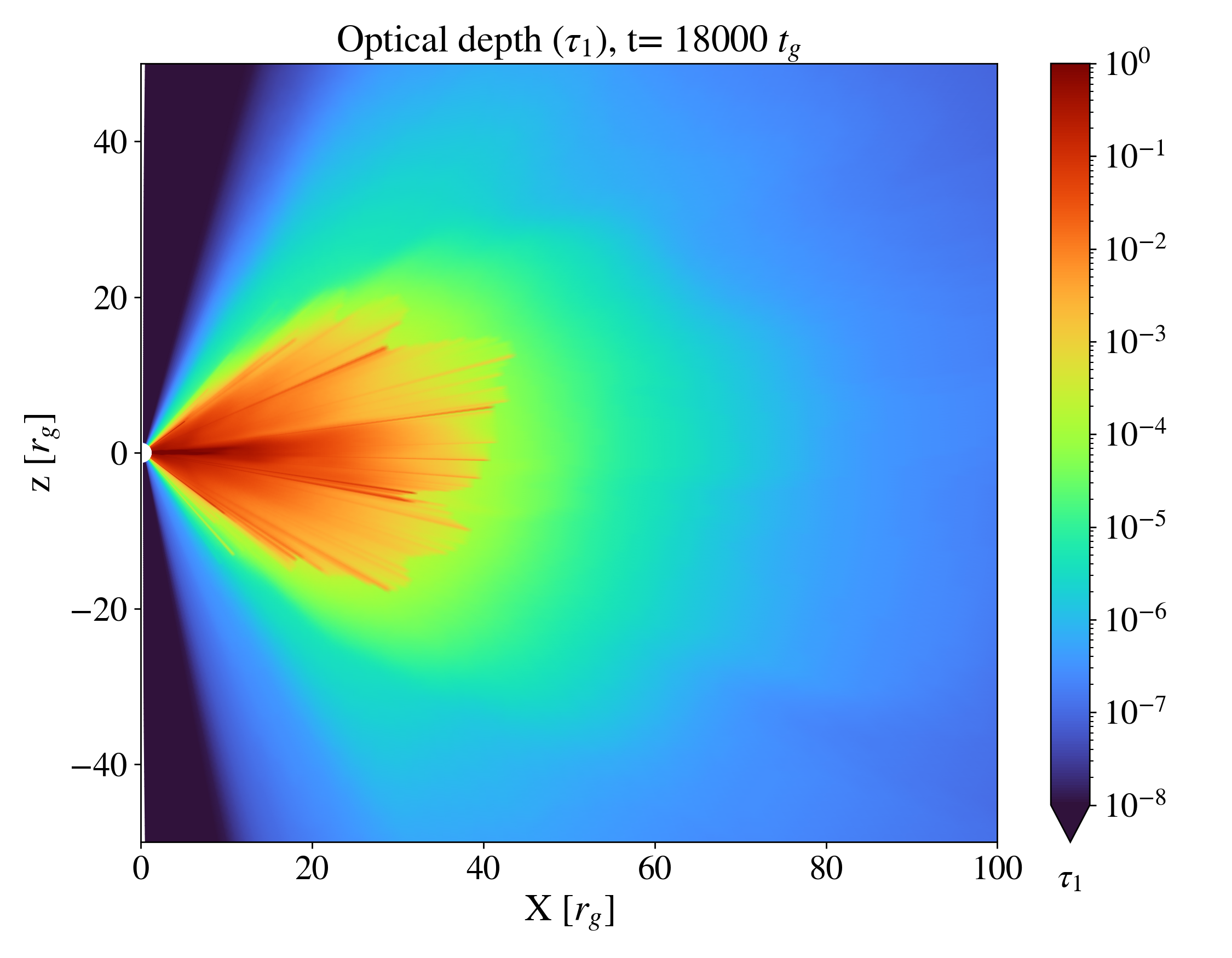}
    \caption{Snapshots from the 2D simulation of the SGRB central engine, as shown in Fig.\ref{fig:GRrun}. The top panel shows the neutrino cooling rate in the innermost region of the accretion disk. The bottom panel shows the color map of the neutrino optical depth.
    }
    \label{fig:Neutrino}
  \end{figure}

Figure \ref{fig:GRrun} shows a snapshot of a fiducial 2D simulation. 
The model shown is 2D-Thin
, and the maps correspond to time $t=18000 t_{g}$.
The polar plots display the density distribution (in code units; top panel) with overplotted magnetic-field lines. 
Maps of additional quantities from the same simulation are presented in Figure \ref{fig:Neutrino}.

Dense, nearly completely neutronized material is located near the equatorial plane of the disk. At intermediate latitudes, the disk launches fast, magnetically driven, and neutrino-cooled outflows. We estimate their velocities to be $v/c \sim 0.11-0.23$.
These outflows are launched with a wide range of electron
fractions, $Y_{\rm e} \sim 0.1-0.45$.
The mass-loss rates and thermodynamic properties of the winds are sensitive to engine parameters, specifically the BH spin and
disk magnetization, as previously shown by \cite{Janiuk2019, Nouri2023}.
In general, more strongly magnetized disks produce faster outflows.
The winds eventually contribute to the kilonova signal through the radioactive decay of r-process isotopes, providing a nucleosynthesis site complementary to the dynamical ejecta launched prior to the compact binary merger.

  Figure \ref{fig:Neutrino} shows color maps of neutrino emissivity and optical depth for the same model and snapshot as in Fig. \ref{fig:GRrun}.
  The simulation is designed to produce an optically thin model. Indeed, the $\tau>1$ region is limited to the narrow strip along the equatorial plane, where neutrino emissivity is also largest. Elsewhere, the medium remains optically thin ($\tau<1$) for both scattering and absorptive processes.

Other 2D models qualitatively follow a similar evolutionary pattern. Quantitatively, the thick scenarios produce a much denser torus and lower electron fractions  near the equatorial plane compared to the thin scenarios. Most of this neutron-rich material is accreted with the inflow of matter, whereas the neutronization of unbound winds is similar in both cases. Quantitative differences in the distribution of $Y_{e}$ result in distinct heavy-element abundance patterns after nucleosynthesis, which are presented in the next section. Differences in the neutrino emissivity distribution are such that thick models are brighter in neutrinos in their central regions, whereas thin models are brightest near the disk surface. In the evolved state, the total neutrino luminosities differ by a factor of $\sim$5-8 (see below).

\begin{figure}
\includegraphics[width=0.49\textwidth]{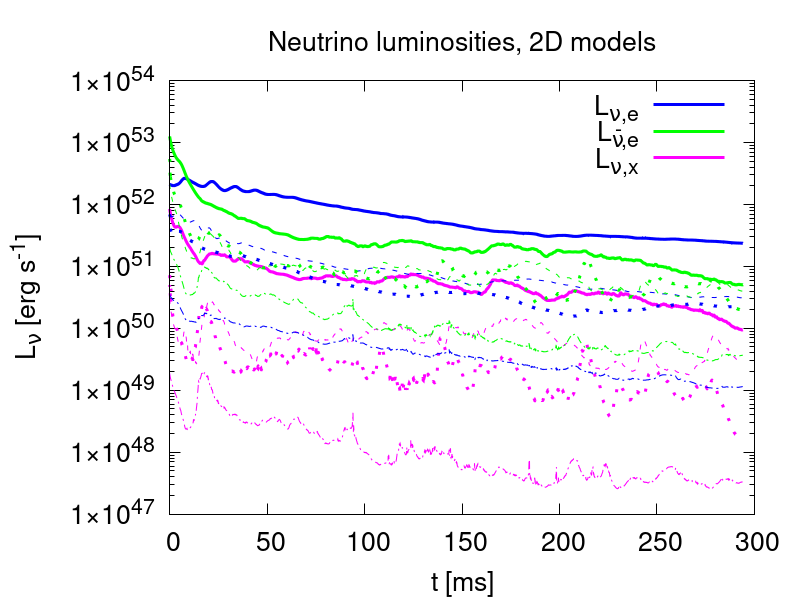}
    \caption{Neutrino light curves for the 2D models with optically thin and thick disks. 
    The plots show the  2D-Thin-s and  2D-Thin models (thin dot-dashed and dashed lines, respectively) and the 2D-Thick-s and 2D-Thick models (thicker dotted and solid lines, respectively). Blue, green, and magenta indicate luminosities in electron neutrinos, electron anti-neutrinos, and heavy-lepton neutrinos, respectively.
    }
    \label{fig:2D_thin_thick_neurinoLC}
  \end{figure}

The neutrino emissivities integrated over the volume give the total neutrino luminosity of the disk and outflows. Figure \ref{fig:2D_thin_thick_neurinoLC} shows this quantity as a function of time for the 2D models. The figure presents three neutrinos flavors. The luminosities of electron and anti-electron neutrinos differ but are both much higher than those of the heavy-lepton neutrinos. In this study, we do not consider neutrino oscillations, so the heavy-lepton neutrino contribution remains energetically negligible (see, however, \cite{Nagakura2025} for recent results on neutrino oscillation effects). 

 We highlight the comparison between the total luminosities of the thin and thick models, including the test models with low disk masses, denoted as 2D-Thin-s and 2D-Thick-s.
 The optically thin models have $L_{\nu} \sim 10^{52}$ erg/s initially, and slowly decrease by approximately one order of magnitude over the 300 ms simulation time. This luminosity correlates with the initial disk mass, which is 0.013 $M_{\odot}$ and is approximately an order of magnitude lower for the test model with a lower disk mass of only 0.002 $M_{\odot}$. These models share the same global parameters (BH spin and magnetic field strength), so we expect that the luminosity of the jet driven by the BH rotation mediated by magnetic fields (i.e., the Blandford-Znajek process) would be the same. 
Since $A_{\phi} \propto \rho$, models with different disk masses may have different magnetic field amplitudes. However, in the code, we renormalize the field at the initial state to obtain the assumed value of minimum $\beta$ of the plasma, which is located close to the inner disk radius, and is used as the model parameter (see Table \ref{tab:Sim_Models}).
However, neutrinos may be a much weaker source of jet power in the low-mass disks (i.e., disk to black hole mass ratio).
At late time of evolution, the slope of the neutrino light curve in the test model flattens, suggesting that the difference in neutrino-power may not be significant at later times.

The thick models are brighter in neutrinos  by more than an order of magnitude, with luminosity $L_{\nu} \le 10^{53}$ erg/s, reflecting the larger disk mass of approximately 0.13 $M_{\odot}$. The initial difference between electron and anti-electron neutrino luminosities arises from larger absorption for $L_{\nu, \bar{e}}$. 
The $L_{\nu, x}$ in the simulation remains an order of magnitude lower. 

The models with low disk mass evolve with a neutrino luminosity slope similar to that of the fiducial models. We speculate that in neutrino thick cases, the jet powered by neutrino annihilation can be launched successfully from the central engine, even though its magnetization is lower and the BZ power is insufficient to sustain it.

 In all 2D models, the accretion disk operates in the SANE (standard and normal) accretion mode. The magnetically arrested disk (MAD) state occurs if the normalized magnetic flux on the BH horizon, $\phi_{BH}= 1/(2\sqrt{\dot m}) \int_{r_{hor}} |B^{r}| \sqrt{-g} d\theta d\phi$, reaches a minimum value of $\sim 15$ (for Gaussian units). In our 2D models, $\phi_{BH}$  never exceeds a value of ten, although it varies overtime in response to accretion rate fluctuations (see below). In all fiducial models,  $\phi_{BH}\sim 1-4$ for most of the simulation time, independent of the dimensionality of the run.

\begin{figure}
\includegraphics[width=0.49\textwidth]{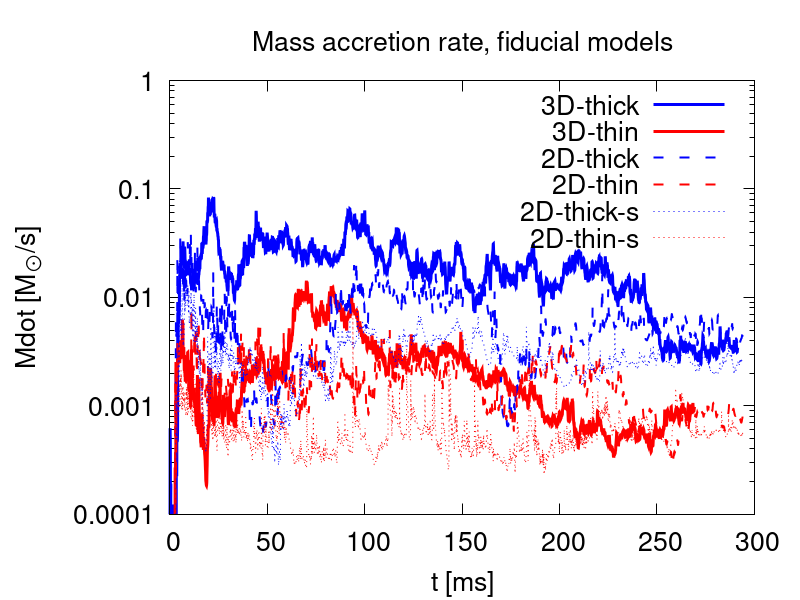}
    \caption{Mass accretion rates for thin and thick models. The 2D-Thin-s, 2D-Thin and 3D-Thin models are shown with dotted, dashed, and solid red lines, respectively. The 2D-Thick-s, 2D-Thick and 3D-Thin models are shown with dotted, dashed, and solid blue lines, respectively. The plot is presented on a logarithmic scale.
    }
    \label{fig:Mdot_in_2D_3D}
  \end{figure}

The effects of 3D simulations on the density distribution are most apparent in the accretion rate. Figure \ref{fig:Mdot_in_2D_3D}  compares the accretion rate at the BH horizon radius as a function of time for optically thin and thick scenarios, modeled in 2D and 3D. The 3D-Thick simulation yields an accretion rate two to three times higher than its 2D counterpart. For the 3D-Thin simulation, the mean accretion rate is lower and at late times it is similar to that in the 2D model. As shown in the following section, the difference is also reflected in the mass outflow rates. We attribute this effect to magnetic turbulence and enhanced angular momentum transport by magnetic fields the full 3D setup, whereas in 2D axisymmetric simulations, angular momentum transport by magnetic fields is numerical rather than physical.

The magnetic flux on the BH horizon in 3D models has values similar to those in the 2D cases. For the 3D-Thin run,  $\phi_{\rm BH}=2-3$ for most simulations and increases to $\sim 5$ toward the end of the run. For the 3D-Thick run, the value is always between $\phi_{\rm BH}=1-2$.

\begin{figure}
\includegraphics[width=0.49\textwidth]{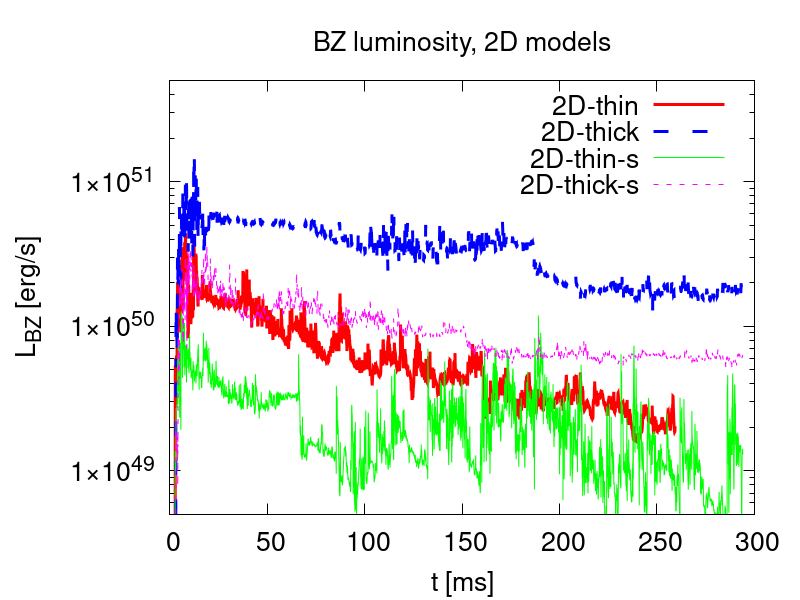}
    \caption{Blandford-Znajek luminosity for thin and thick models. The 2D-Thin-s and 2D-Thin models are shown with solid green and red lines, respectively. The 2D-Thick-s and 2D-Thick models are shown with dashed magenta and blue lines, respectively.
    }
    \label{fig:BZ_2D_thin_thick}
  \end{figure}

Figure\ref{fig:BZ_2D_thin_thick} shows the Blandford-Znajek (BZ) power for the 2D fiducial models. The BZ power is consistently about an order of magnitude smaller than the electron-neutrino luminosity for the corresponding models (cf. Fig. \ref{fig:2D_thin_thick_neurinoLC}). We argue that the BZ mechanism in these simulations provides a weaker source of power for the GRB jets than neutrino annihilation, unless the efficiencies of the processes differ substantially.
  
\subsubsection{Kilonova central engine models} 

In addition to the thin and thick models, we performed simulations with central engine parameters tuned to represent kilonova sources. In particular, we tested a scenario for GRB 211211A. The engine parameters reflect the compact binary mass ratio and the final BH spin after the merger. The 2D-kn, 3D-kn and 3D-kn-p models were run 
until a final time of 20,000 $t_{g}$. In one of the 3D models, we introduced a non-axisymmetric perturbation of the internal energy at initialization. We aimed to reproduce the physical properties of the system, prolong the activity time, and examine the implications for heavy-element nucleosynthesis. In this way, our objective was to verify whether a compact binary merger scenario is plausible for this source, alongside other ideas recently proposed in the literature.

\begin{figure}
   \includegraphics[width=0.49\textwidth]{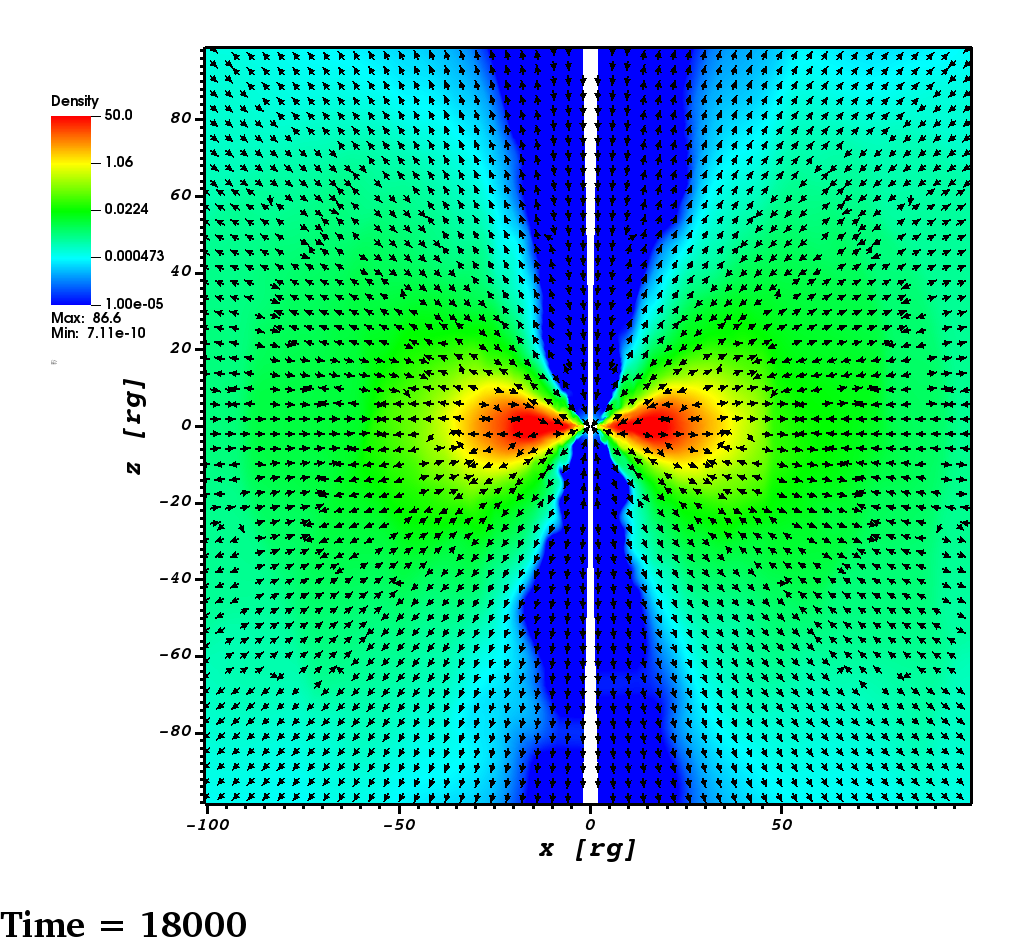}
\includegraphics[width=0.49\textwidth]{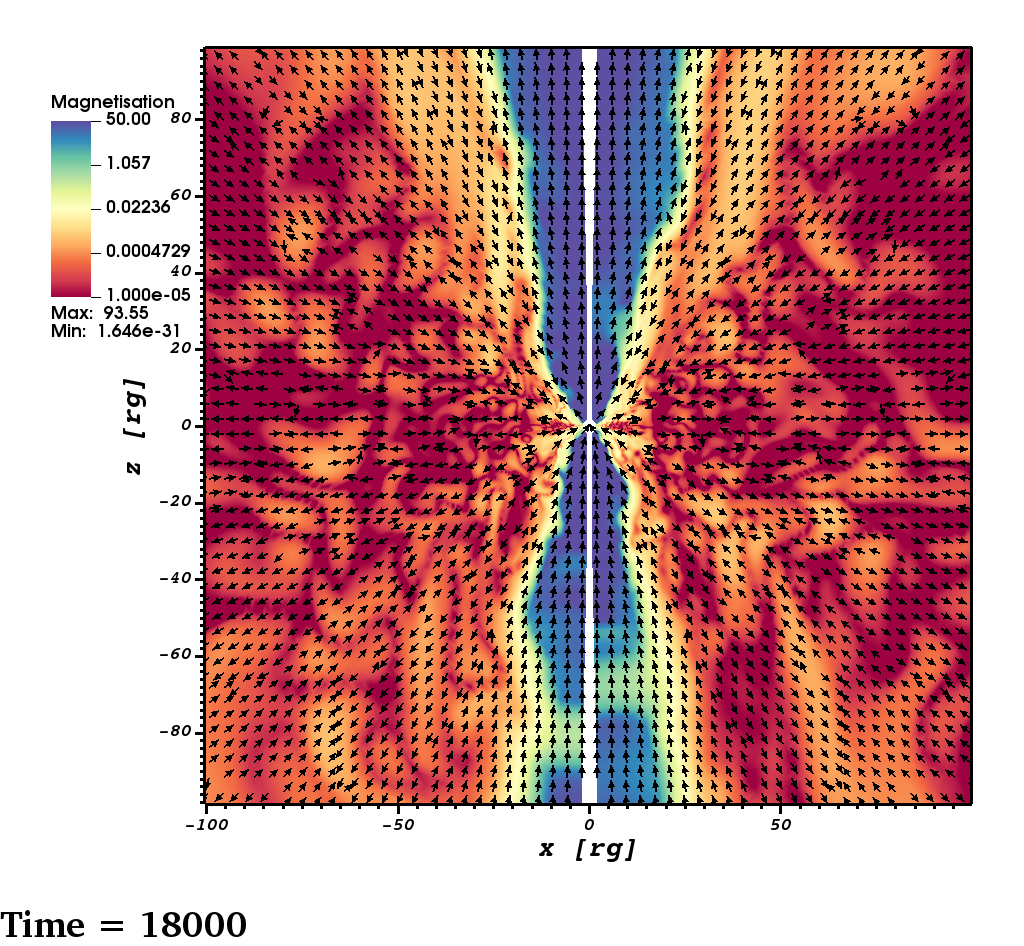}
    \caption{Distributions of density $\rho$ (in code units) with superimposed velocity field, and magnetization $\sigma=B^{2}/\rho$ with magnetic field vector fields, at time t=18000 $t_{g}$, for the 3D kilonova model, 3D-kn. 
    }
    \label{fig:Density_KN3D}
  \end{figure}

\begin{figure}
   \includegraphics[width=0.49\textwidth]{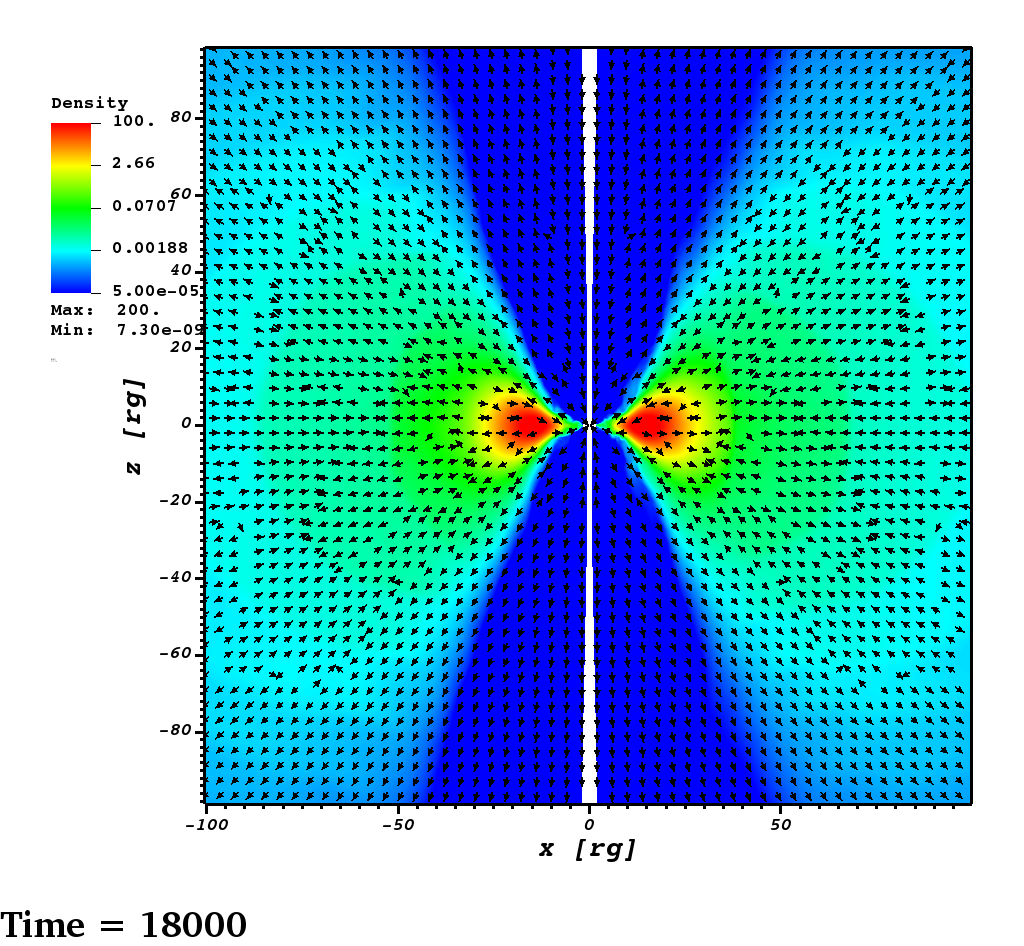}
\includegraphics[width=0.49\textwidth]{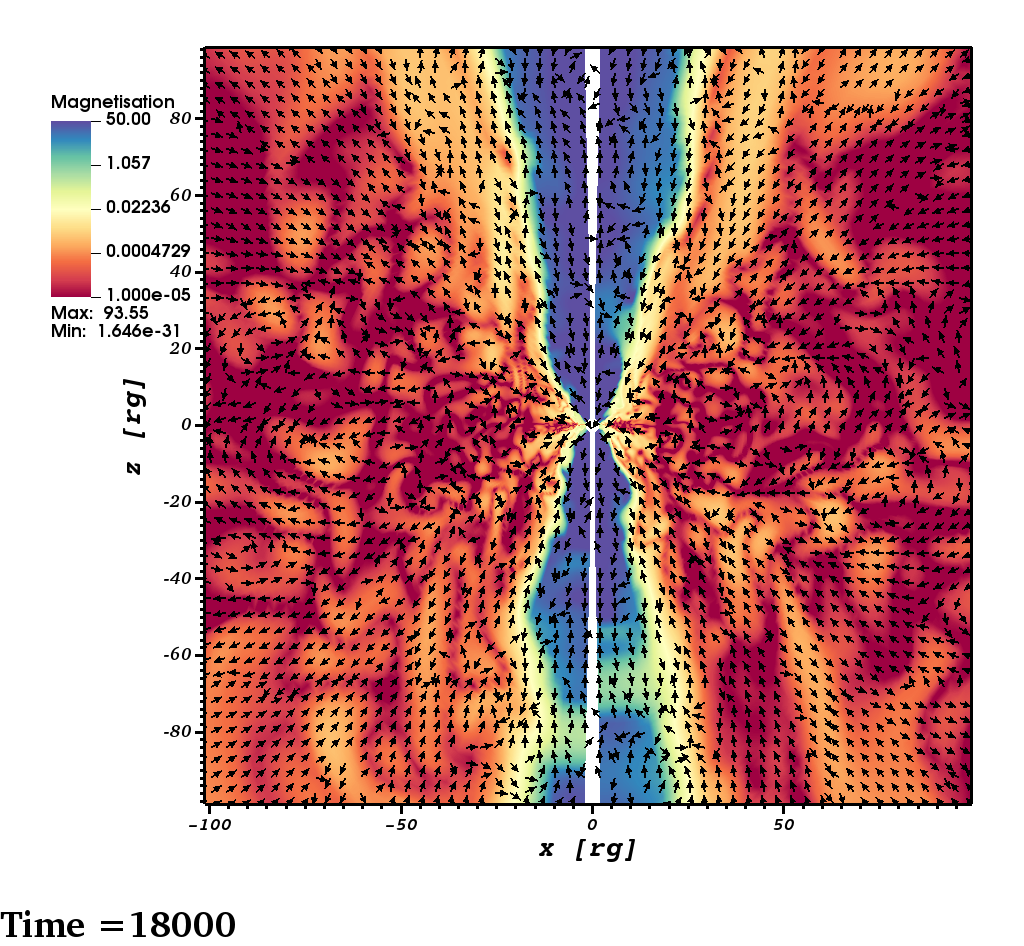}
    \caption{Distributions of density $\rho$ (in code units) with superimposed velocity field, and magnetization, $\sigma=B^{2}/\rho$ with magnetic field vector fields, at time t=18000 $t_{g}$,for the 3D kilonova model with non-axisymmetric perturbation, 3D-kn-p 
    }
    \label{fig:Sigma_KN3D}
  \end{figure}

Figures \ref{fig:Density_KN3D} and \ref{fig:Sigma_KN3D}present polar slices of the 3D models.
The top panels show the density and velocity field distribution, while the bottom panels show magnetization in the flow.
Figure \ref{fig:Density_KN3D} shows the 3D-kn model initialized in an axisymmetric setup, whereas Fig \ref{fig:Sigma_KN3D} shows scalar and vector fields for the 3D-kn-p model, with the initial perturbation.
Quantitative differences are apparent between the two models at the evolved state of $t=18,000 ~t_{g}$.In the perturbed model, the low-density funnel becomes wider, corresponding to a potentially larger jet opening angle. At the same time, the accreting torus in the 3D-kn-p model is denser and more compact within the innermost $\sim 30 ~r_{g}$. The magnetization profiles reveal distinct shapes of turbulent regions within the torus and in the surface layers where the winds are launched. The winds from the perturbed model are less dense and distributed more towards the equatorial plane than the winds from the unperturbed scenario.
The mass loss and other properties of the winds, which influence subsequent nucleosynthesis patterns, also differ; these differences are discussed in the next section.

\begin{figure}
\includegraphics[width=0.49\textwidth]{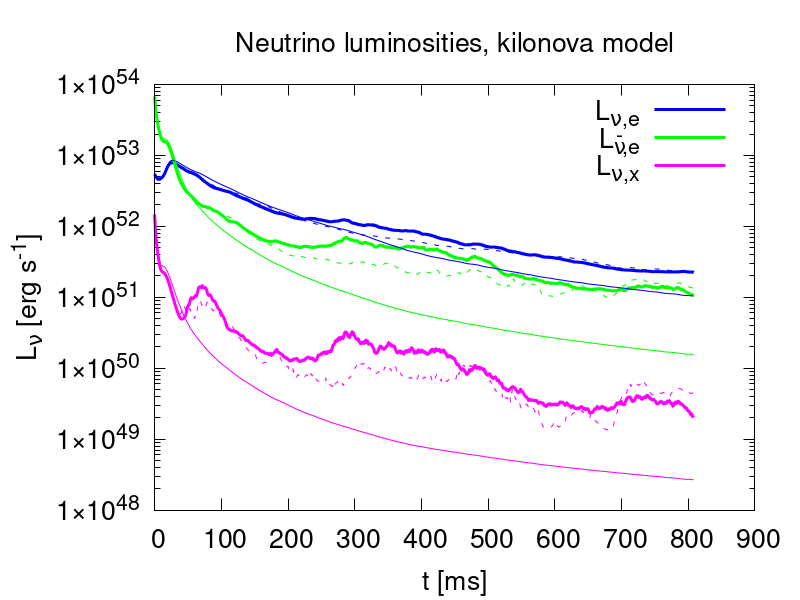}
    \caption{Neutrino light curves for the kilonova scenario. The 2D-kn and 3D-kn models are shown with dashed and solid lines, respectively. The results for the 3D-kn-p model are shown with a thinner solid line. The line colors represent the various neutrino flavors, as in Fig. \ref{fig:2D_thin_thick_neurinoLC}.}
    \label{fig:2D_kn_neurinoLC}
  \end{figure}

We note that the 2D-kn and 3D-kn models are quantitatively similar in torus structure, mass loss, and neutrino luminosity, as shown in Fig. \ref{fig:2D_kn_neurinoLC}. In the figure, the dashed and thick solid lines almost completely overlap within the first $\sim 300$ ms for all neutrino flavors.
The azimuthally perturbed model, 3D-kn-p exhibits a lower neutrino luminosity than the axisymmetric model,  and it declines at a somewhat faster rate.
The average electron neutrino luminosity obtained in our simulations is $\sim2.3\times 10^{51}$ erg s$^{-1}$ for the 3D-kn model and $1.1\times 10^{51}$ erg s$^{-1}$for the 3D-kn-p  model.
Considering that recent estimates place the neutrino annihilation efficiency at $\sim 0.1-1\%$ \citep{2019MNRAS.482.2973D},  we speculate that the low $L_{\nu}$ implies that this engine cannot power a bright GRB jet for an extended period.
  
\begin{figure}
\includegraphics[width=0.49\textwidth]{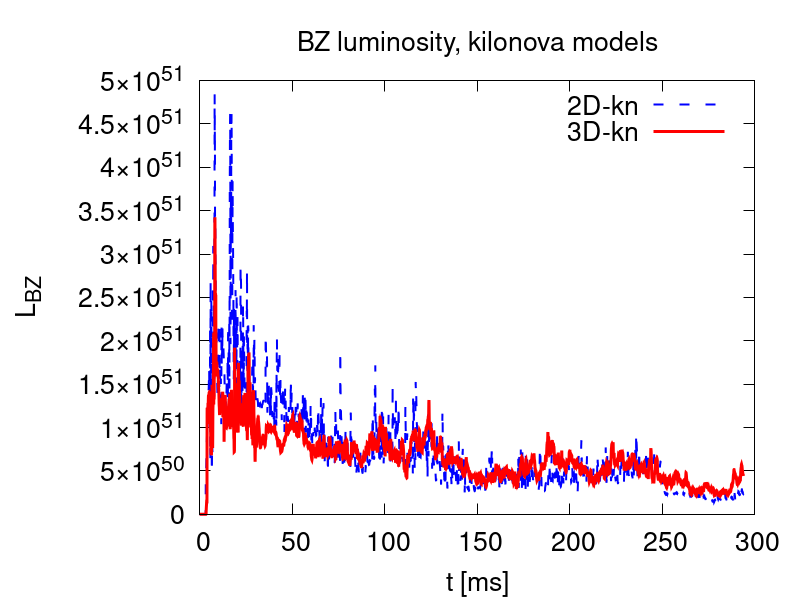}
    \caption{Blandford-Znajek light curves for the kilonova scenario. The 2D-kn and  3D-kn models are shown with solid blue and red lines, respectively.} 
    \label{fig:2D_3D_kn_LBZ}
  \end{figure}

\begin{figure}
\includegraphics[width=0.49\textwidth]{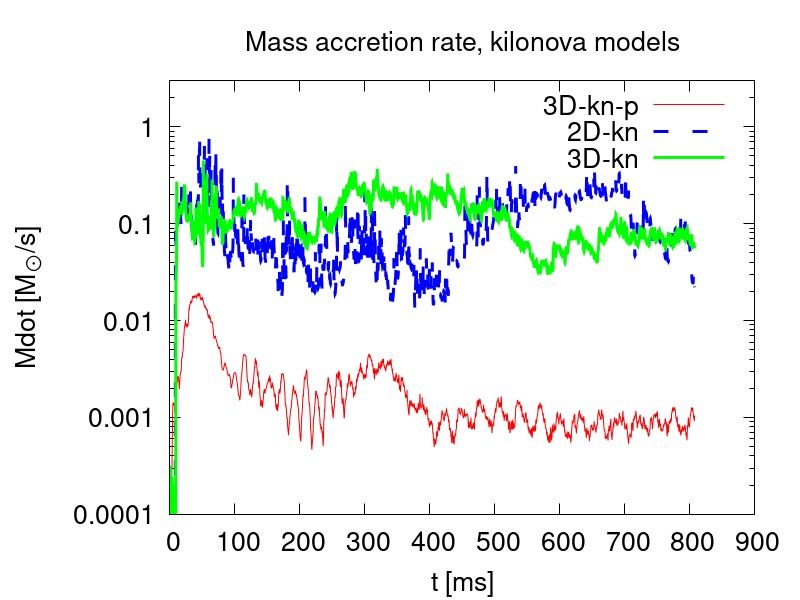}
    \caption{Mass accretion rate in the kilonova scenario. The 2D-kn and 3D-kn models are shown with blue and green lines, respectively. In addition, the 3D-kn-p model is shown with a thinner red line.  The plot is shown on a logarithmic scale} 
    \label{fig:2D_3D_kn_mdot}
  \end{figure}

Figure \ref{fig:2D_3D_kn_mdot} shows the mass accretion rate at the BH horizon, for the 2D-kn, 3D-kn, and 3D-kn-p models. The accretion rates in 2D and 3D simulations are comparably large and 
are about an order of magnitude larger than those in the fiducial models presented above. As a result, the engine lifetime is longer and accreting tori can power the GRB and kilonova over extended timescales, launching powerful winds.
The non-axisymmetrically perturbed model,  3D-kn-p, however, shows a negligible accretion rate for most of the simulation. The magnetic field is also not transported towards the horizon, and, as discussed above, the neutrino luminosity of this model is also low. 
The simulated time span is short, $\sim 1$~s compared to the timescales of a SGRB. The model does not show magnetic flux accumulation on the BH horizon, and the dimensionless magnetic flux remains at$\sim 2.5$, with no systematic increase over time.  If additional magnetic flux is supplied to the disk at later times, jet launching could occur over longer timescales.
We
 speculate, however, that in this scenario the jet  
 may not have a sufficient power source, either from neutrinos or from the BZ mechanism. The system
 may therefore result in a failed jet, with possible orphan kilonova emission from unbound winds.
Fig. \ref{fig:2D_3D_kn_LBZ} shows the BZ luminosity of the 2D-kn and 3D-kn models.
Both simulations show a similar order of magnitude in BZ power, which is about two orders of magnitude smaller than the corresponding electron neutrino luminosities. The dimensionality affects the temporal variability of the luminosity, which is less pronounced in the 3D setup.

In the next section, we quantify the properties of these winds and the resulting nucleosynthetic patterns in all our scenarios.

\subsection{Mass loss in the outflows from the accretion disk}

\begin{figure}
\includegraphics[width=0.49\textwidth]{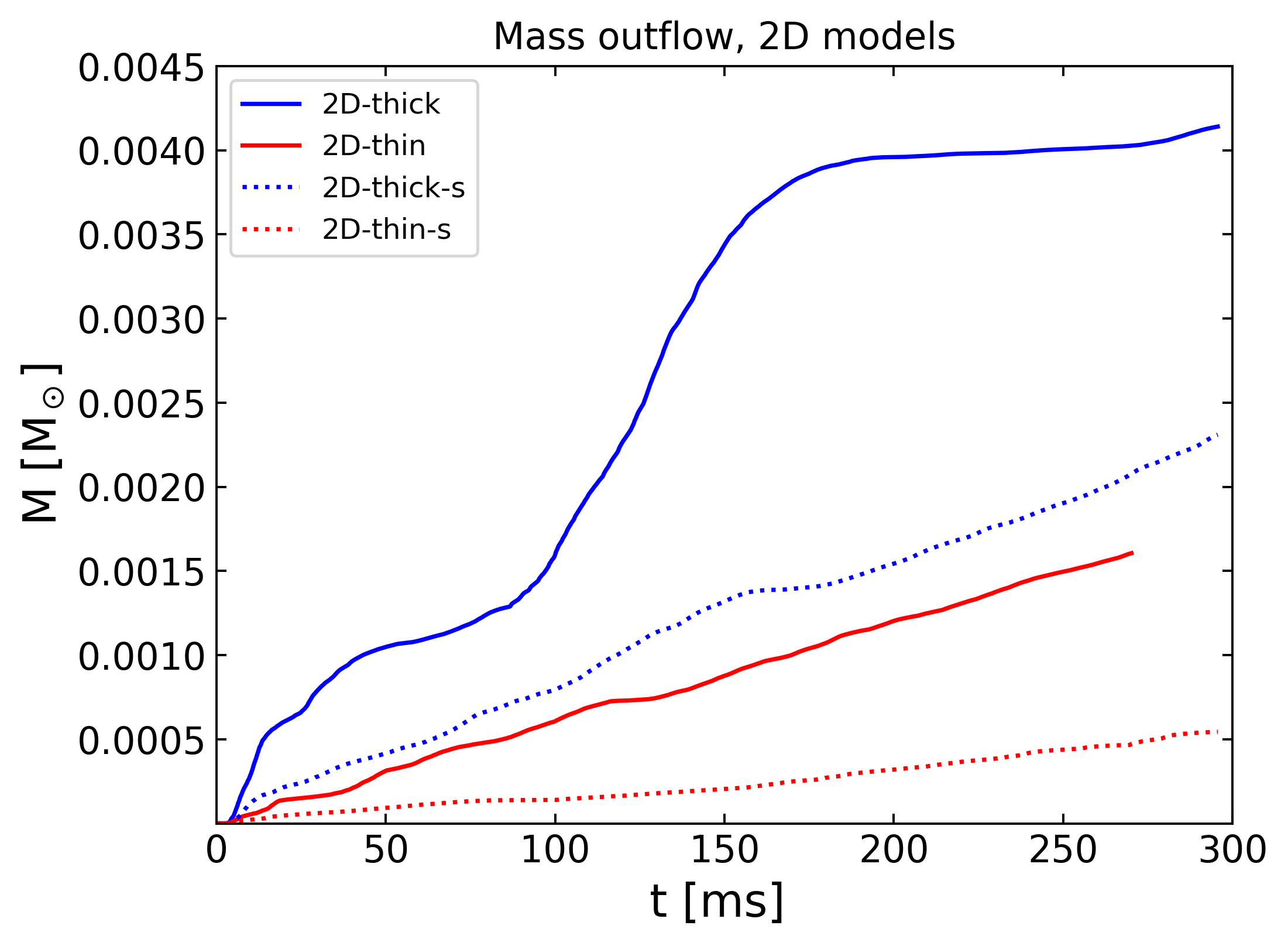}
    \caption{Cumulative mass outflow from the outer boundary for the 2D models.
    }
    \label{fig:2D_mass_out}
  \end{figure}

In general, outflows are launched in all simulations; however, their properties differ. To quantify the outflows, we first measure the mass flux integrated over the surface of the outer boundary. This quantity gives the cumulative mass of the outflows as a function of time. Figure \ref{fig:2D_mass_out} shows the results for the 2D simulations. 
The largest outflow mass, $\sim 0.0045$ $M_{\odot}$, is produced by the 2D-Thick model, which has the largest initial mass. However, the ratio between the ejected mass and the initial disk mass is only about 3.5\%. The 2D-Thin model, as well as the 2D-Thin-s and 2D-Thick-s models, do not reach the state in which the mass outflow saturates at a constant value. The outflow mass initially increases linearly and is expected to saturate later, once the simulation reaches a steady state. Nevertheless, by the end of this simulation, the torus in the 2D-Thin model has lost about 11\% of its initial mass (0.0015 vs. 0.013 $M_{\odot}$). The optically thin model
is therefore more efficient at launching winds from the accretion disk.

\begin{figure}
\includegraphics[width=0.49\textwidth]{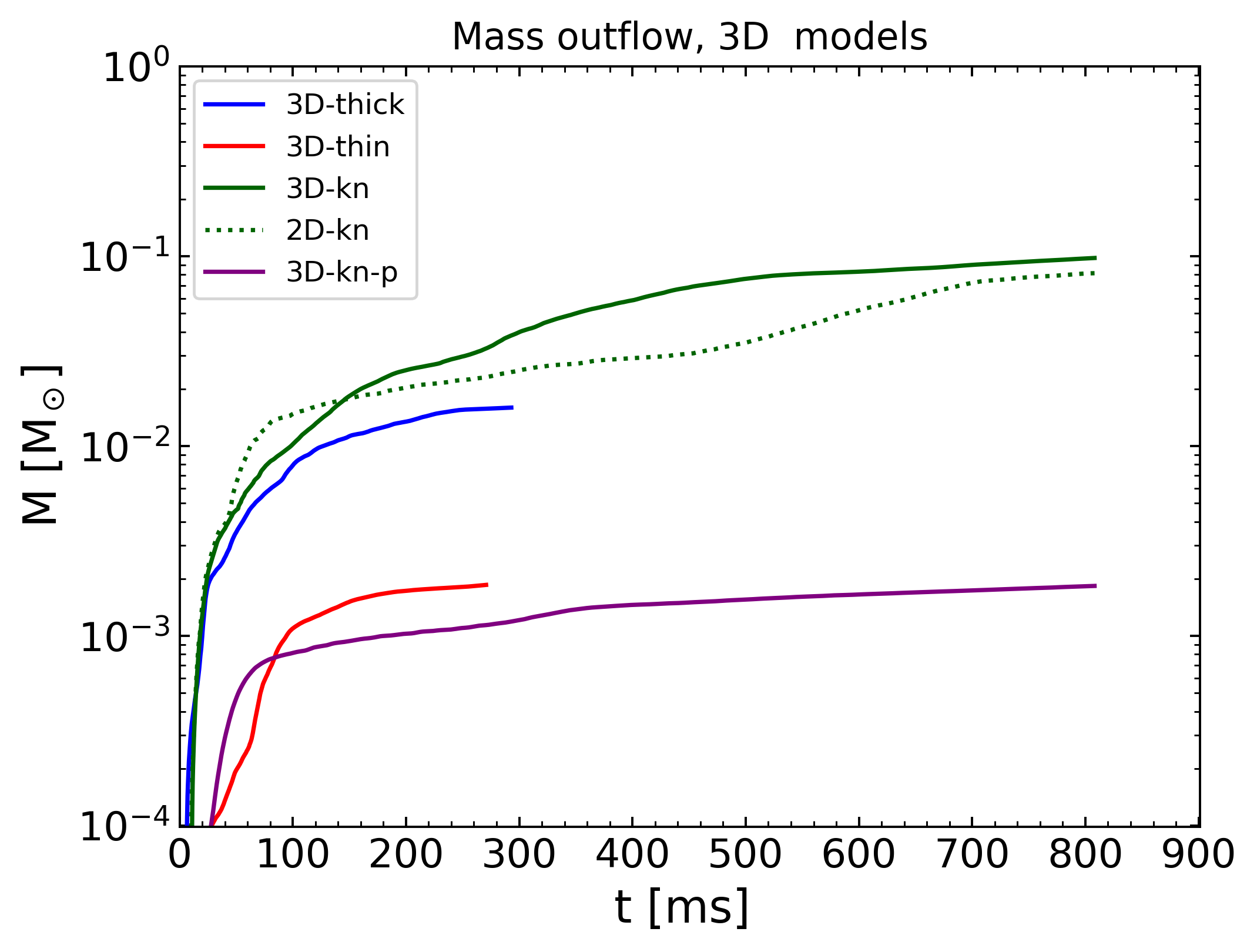}
    \caption{Cumulative mass outflow from the outer boundary for the 3D models. The plot uses a logarithmic scale 
    The time expressed in physical units differs between the fiducial and ``kn'' models because of different $t_{g}$ values, while all models were evolved over the same duration in code units.
    }
    \label{fig:3D_mass_out}
  \end{figure}

Next, we examine whether the mass outflow rate depends on the dimensionality of the simulation. Figure \ref{fig:3D_mass_out} compares the mass outflow rates from the 3D-kn and 2D-kn models. The two cases are very similar, both reaching about 0.1 $M_{\odot}$ at the end of the simulations. This corresponds to roughly 12\% of the initial torus mass. The dimensionality does not affect the results, consistent with what was already indicated by the neutrino luminosities. 
However, for the perturbed model, the mass outflow rate is nearly 100 times lower than in the unperturbed model, for the same initial disk mass. It is comparable to the low-mass 3D-kn model, which is axisymmetric, but has an initial torus mass lower by more than a factor of 50. This demonstrates that in perturbed models
the flow is more bound. We further quantify our estimations via tracer particle analysis in Sect. \ref{sect:tracers}.

\begin{figure}
    \includegraphics[width=0.49\textwidth]{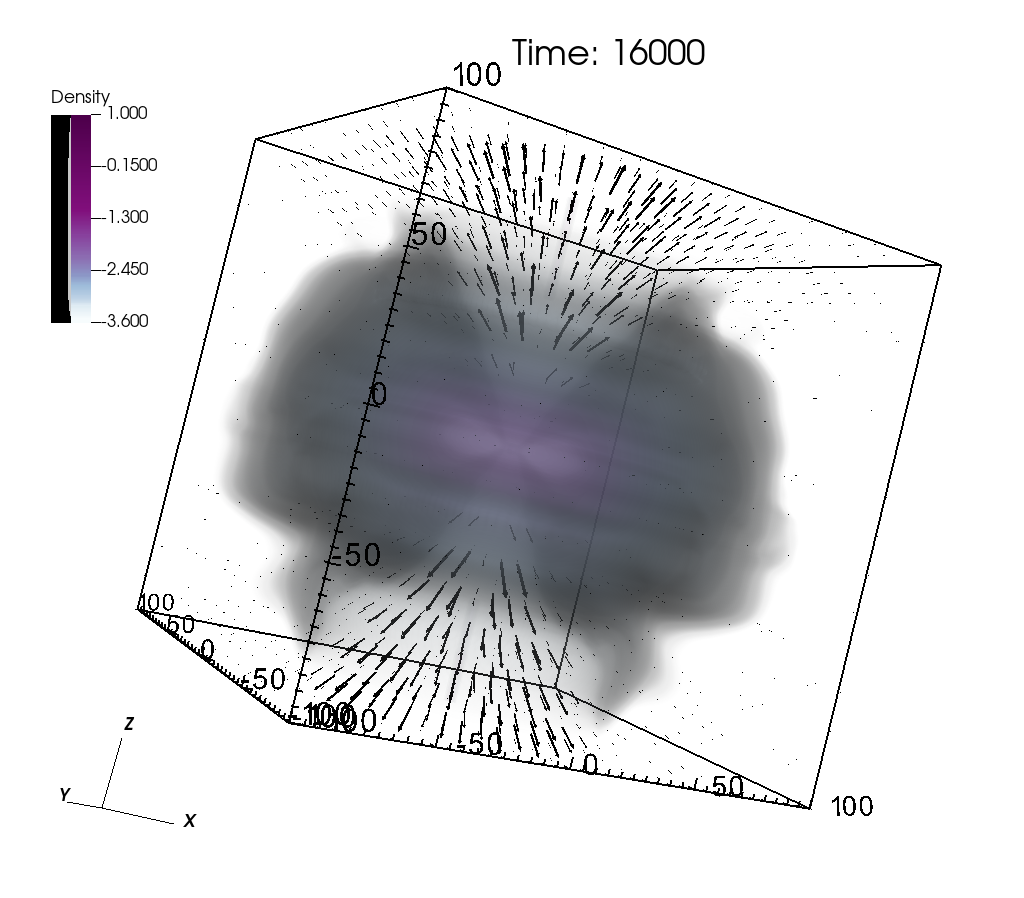}
    \caption{Volume rendering plot of the density with superimposed velocity field in the wind outflow. The plot shows axisymmetric kilonova model  3D-kn at the final time snapshot, $t=16,000 t_{g}$. The density is plotted on a logarithmic scale, in code units.}
   \label{fig:3D_rendering}
  \end{figure}

Figure \ref{fig:3D_rendering} shows a visualization of the 3D axisymmetric kilonova engine, model 3D-kn. The snapshot shows the distribution of the density within the inner 100 $r_{g}$ region and the corresponding velocity field. A powerful wind outflow is clearly launched from the accretion disk.

\subsection{Evolution of outflows probed by tracer particles}
\label{sect:tracers}

\begin{figure}
  \centering
\includegraphics[width=0.2\textwidth]{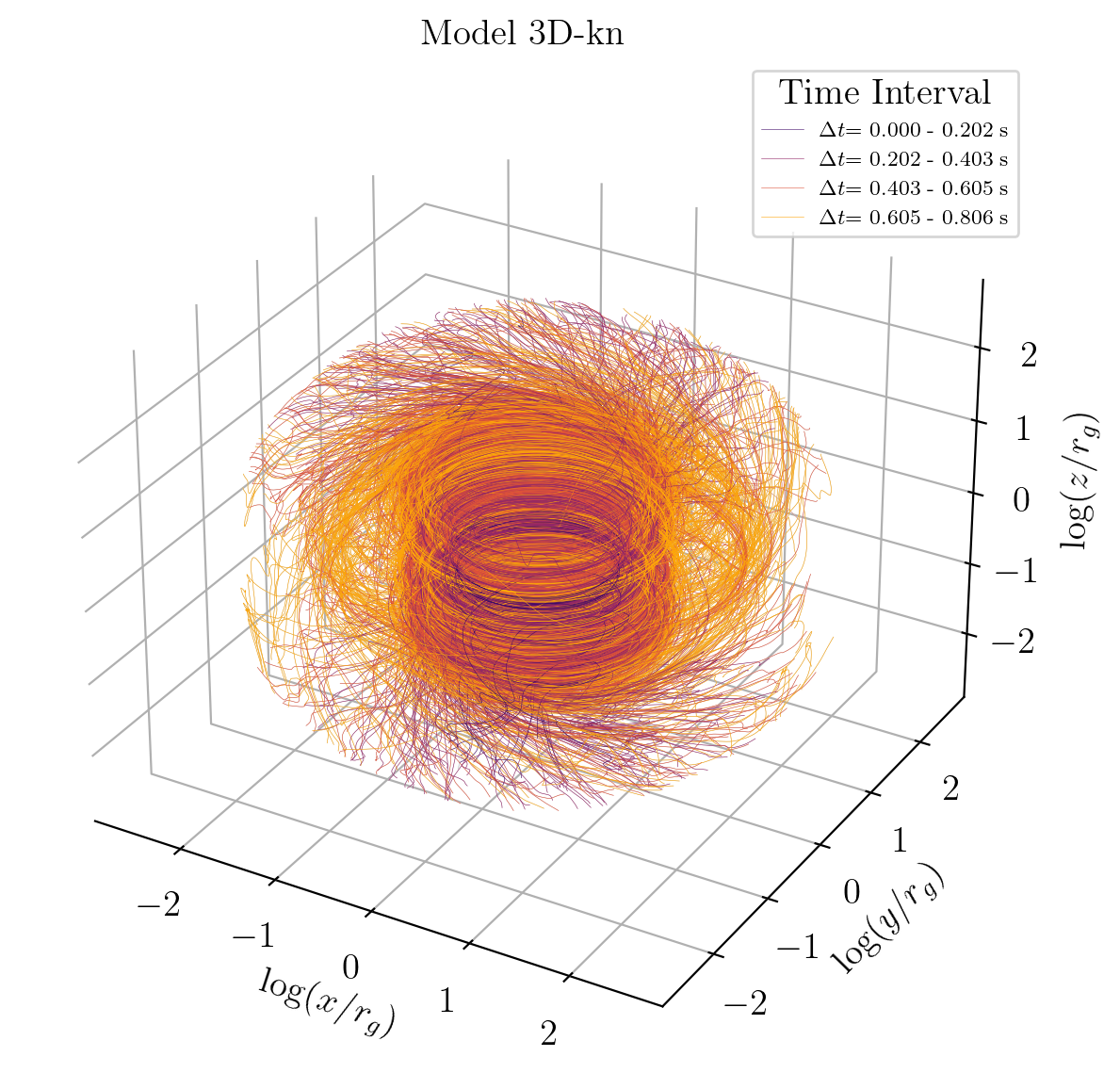}
\includegraphics[width=0.2\textwidth]{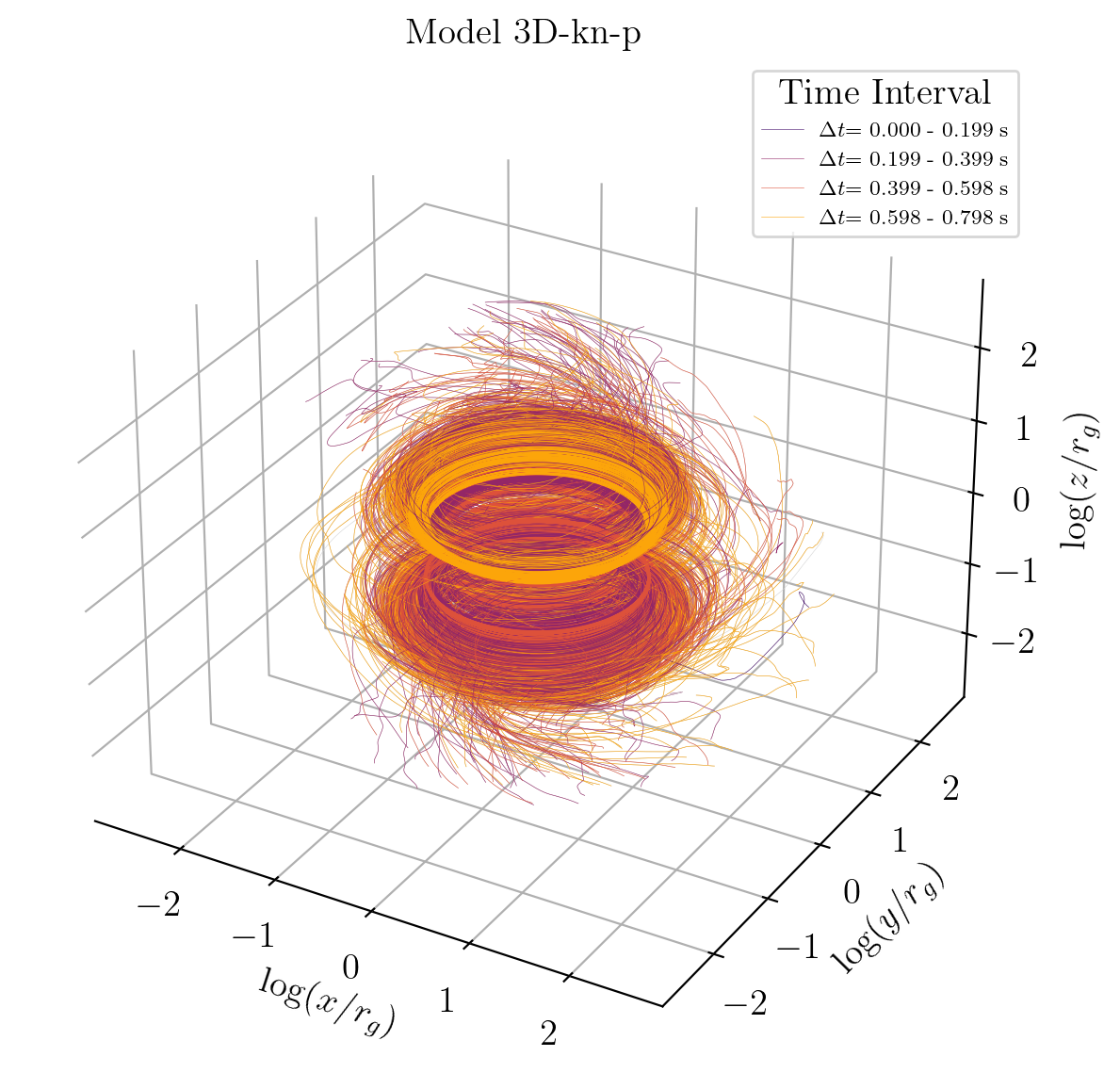}
\includegraphics[width=0.2\textwidth]{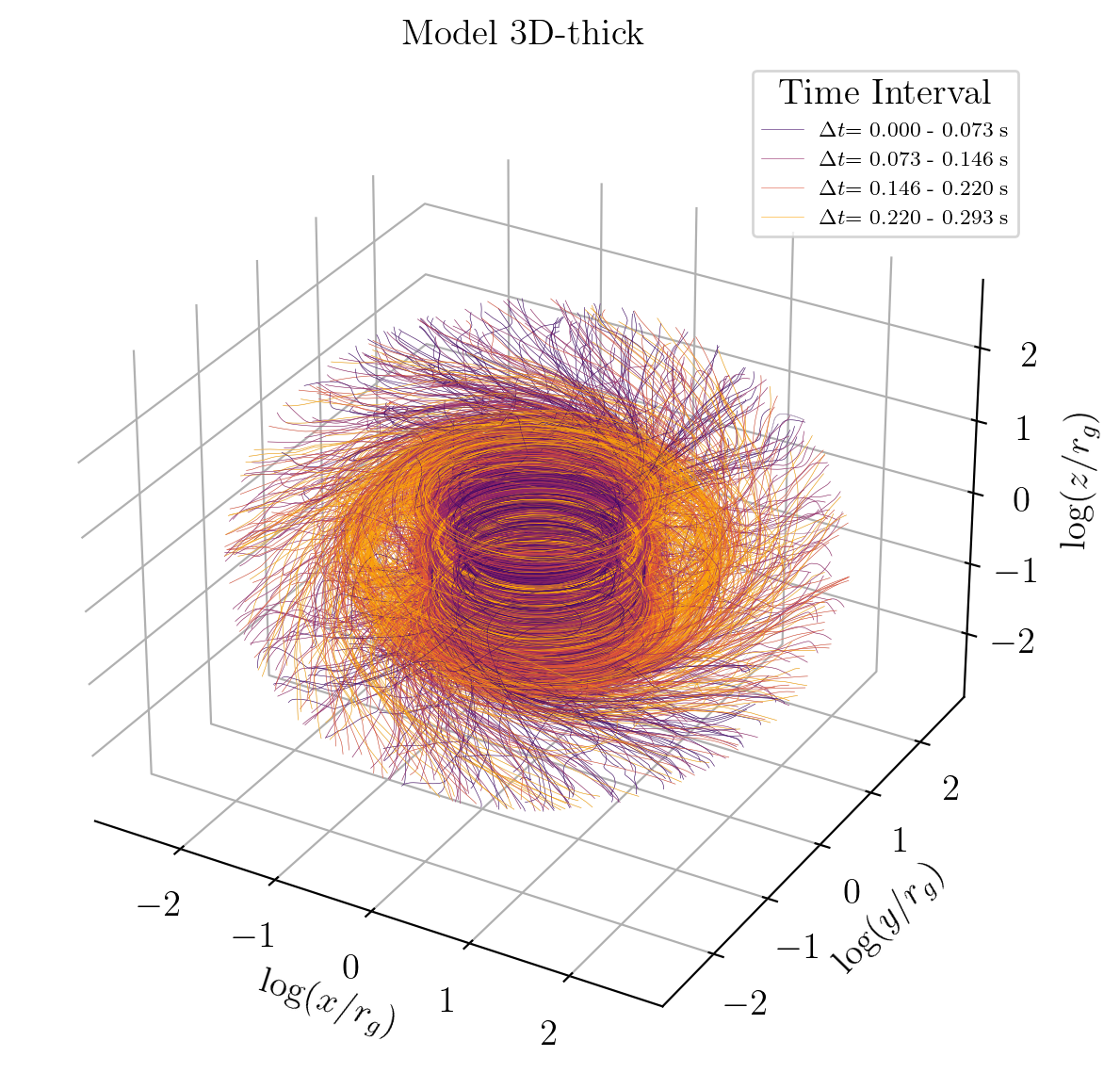}
\includegraphics[width=0.2\textwidth]{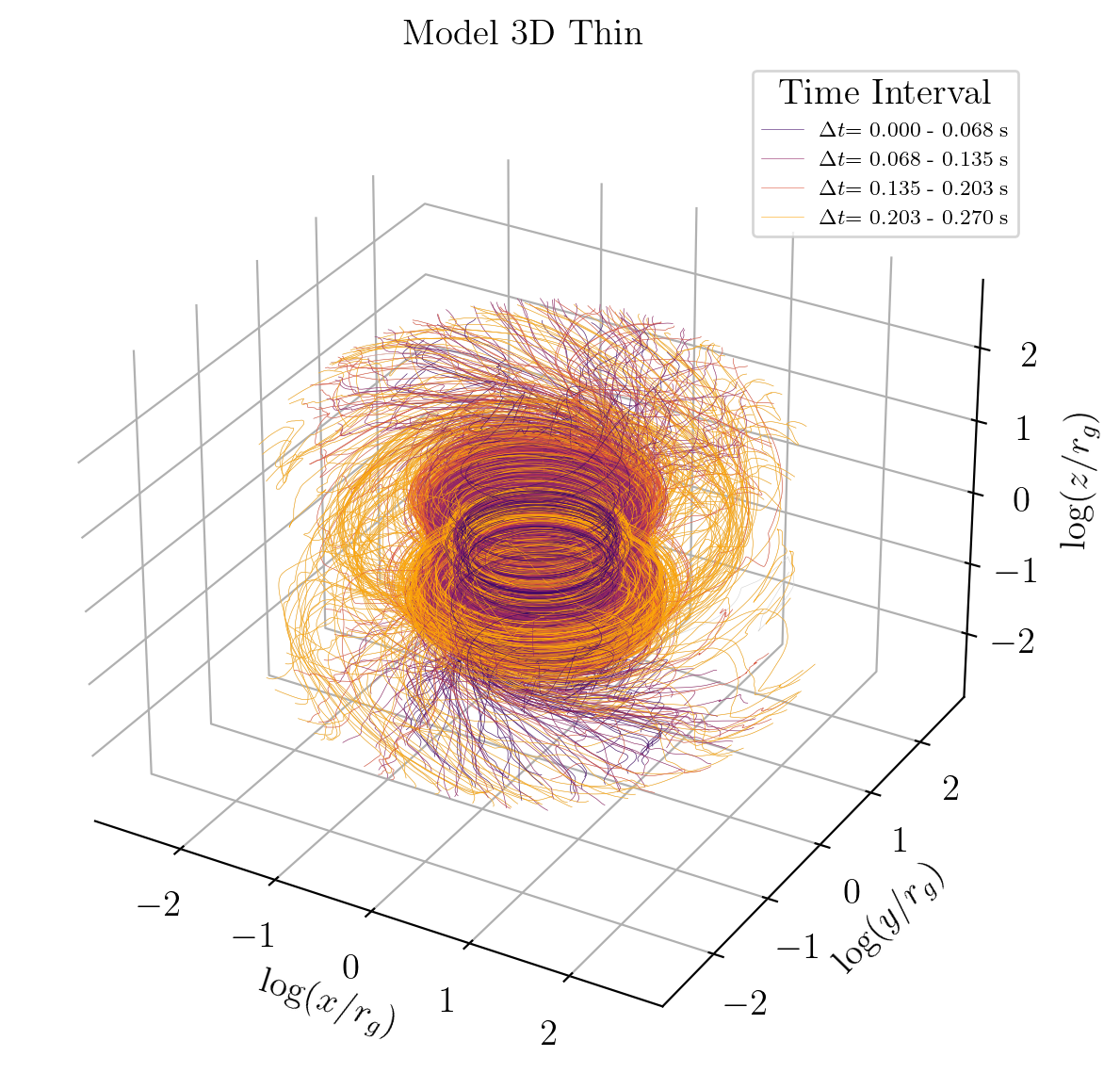}
\caption{Distributions of particle trajectories sampling the outflow from the central engine. The top panels show the
axisymmetric and perturbed kilonova models, while bottom panels show the fiducial thin and thick models. In the 3D-Thin model, the cumulative ejected mass is quantitatively similar to that in the perturbed 3D-kn-p model. The same holds for 3D-Thick and 3D-kn models.} 
    \label{fig:Tracers}
  \end{figure}

To analyze disk outflows, we employed the tracer particle technique originally introduced by \cite{Janiuk2019} for the HARM\_COOL code. 
The tracer particles record properties of the flow, such as density, temperature, and electron fraction, in the disk winds that escape through the outer boundary of the computational grid.
Figure \ref{fig:Tracers} visualizes the subset of particles ejected in all our 3D simulations. Each panel shows a 3D Cartesian box with coordinates scaled logarithmically and normalized in units of the gravitational radius, $r_g$. Each color represents a specific time interval
during which the outflow parcel reaches the outer boundary.
The top panels show the 3D-kn and 3D-kn-p models. The trajectories in the perturbed model mostly cycle around the BH, while only a small subset of particles escapes the computational box. However, color coding reveals that these parcels have high velocities and escape at early time intervals.
The bottom panels show the thin and thick models, 3D-Thin and 3D-Thick. For  the 3D-Thin model, although more particles in total leave the box as unbound winds, their arrival at the outer boundary occurs later. These particles are slower and more massive. Consequently, even though the resulting cumulative mass loss in the 3D-Thin model is very similar to that in the 3D-kn-p case (Fig. \ref{fig:3D_mass_out}), the wind properties and their time histories are different. The chemical properties may also differ between our fiducial models and the kilonova scenario (see Sect. \ref{sect:nuclear}).
For the 3D-Thick model, the outflowing particles move faster and reach the outer boundary earlier, whereas the cumulative outflow mass is an order of magnitude larger than in the previously discussed models and comparable to that of the 3D-kn case. The latter likely contains heavier particles in the unbound material, whose composition is discussed in Sect. \ref{sect:nuclear}).
The cumulative mass outflow is greater by a factor of a few because the initial disk mass is six times larger in 3D-kn than in 3D-Thick.

\subsection{Chemical composition and nuclear reaction network calculations}
\label{sect:nuclear}

We computed the r-process nucleosynthesis in the GRB outflows using tracer data post processed with the nuclear reaction network. We used the SkyNet code \citep{Lippuner_2017} to simulate nuclear reactions and evolve the abundance of nuclear species synthesized in these 
reactions. SkyNet incorporates both strong and weak reaction libraries, allowing the evolution of a wide range of nuclear processes, including symmetric and spontaneous fission. The energy released in the reactions was used to update the temperature self-consistently, while Coulomb screening corrections were excluded in our setup.

 \begin{figure}

 \includegraphics[width=0.24\textwidth]{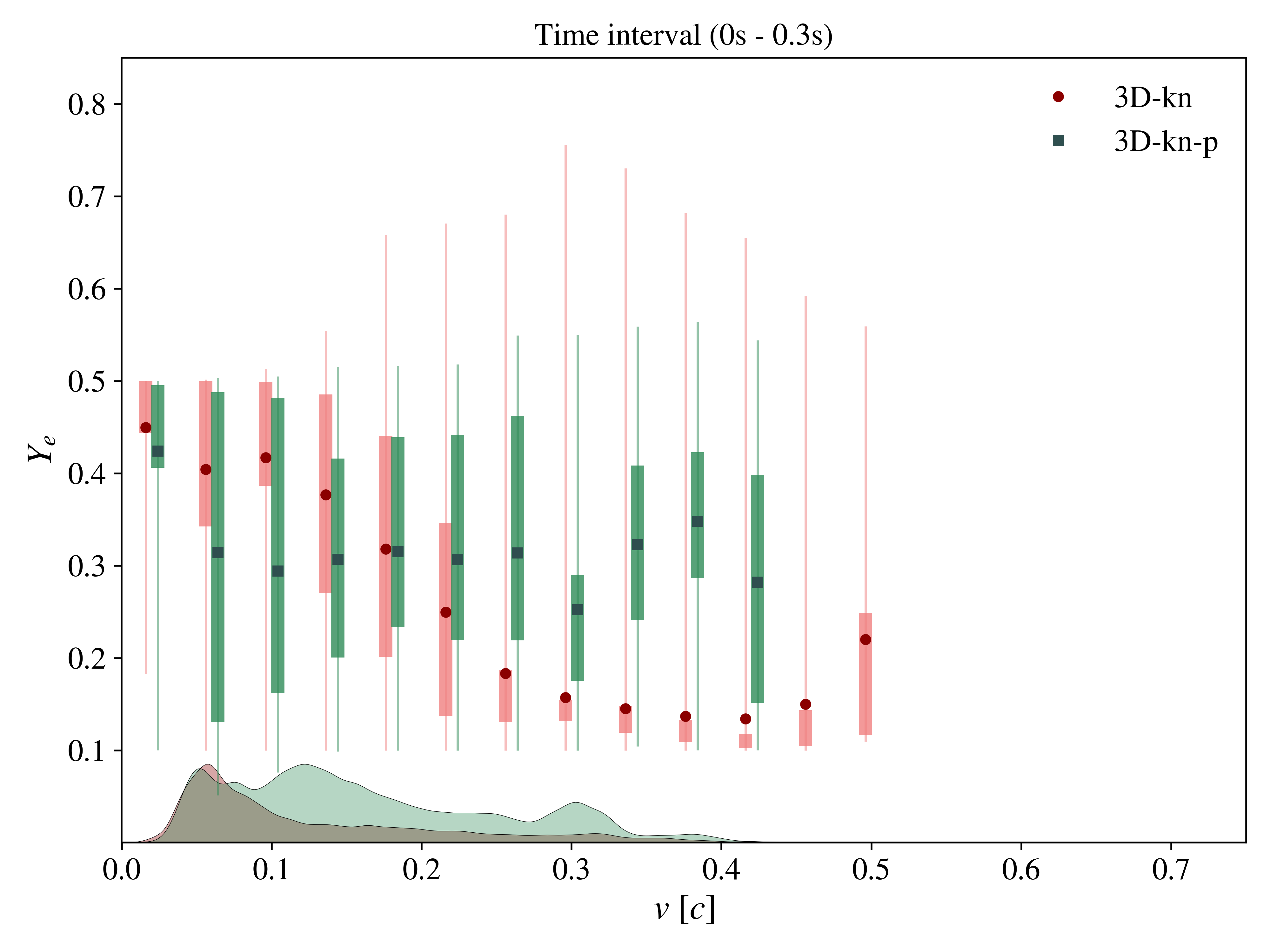}
\includegraphics[width=0.24\textwidth]{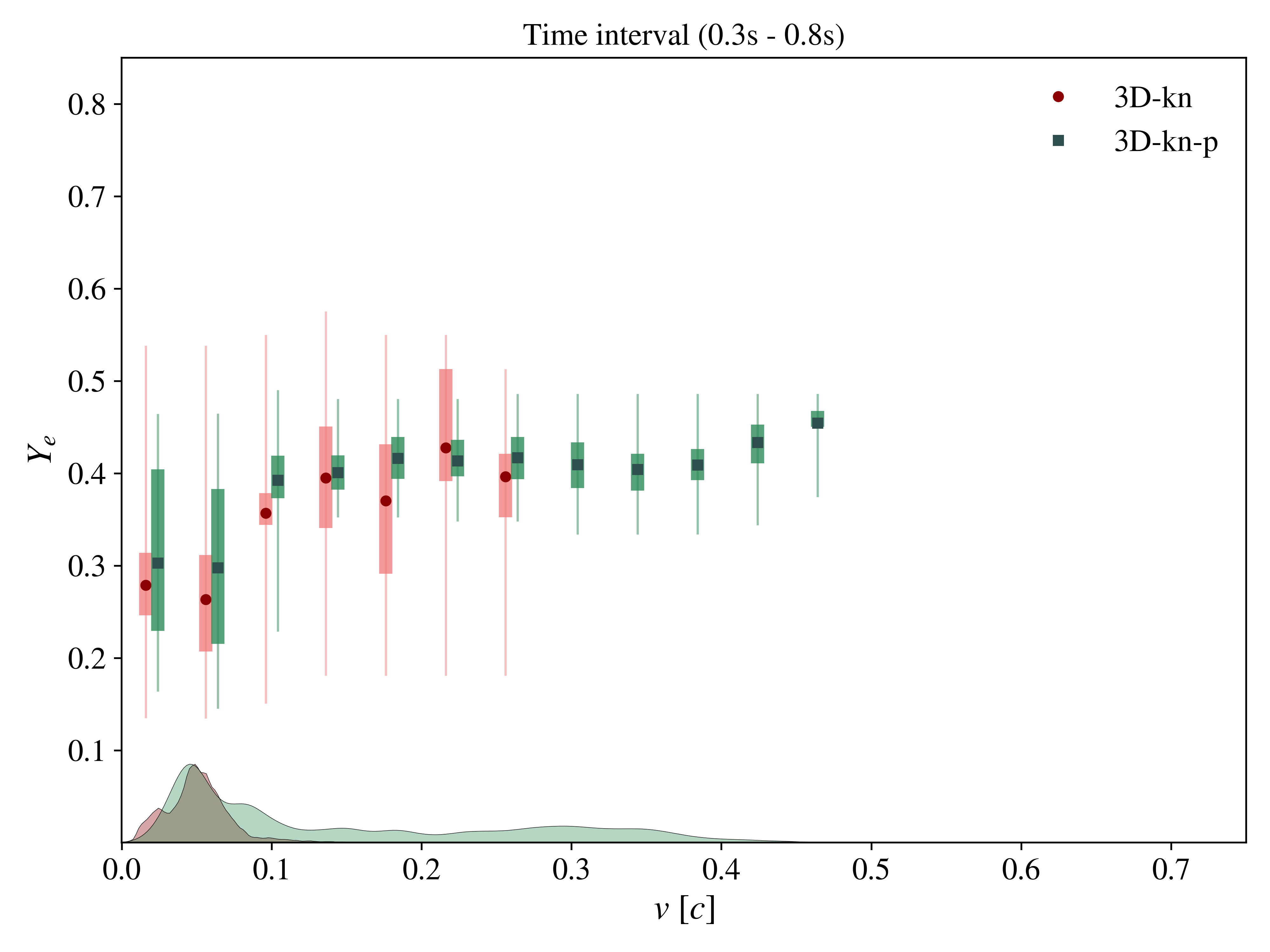}

\includegraphics[width=0.24\textwidth]{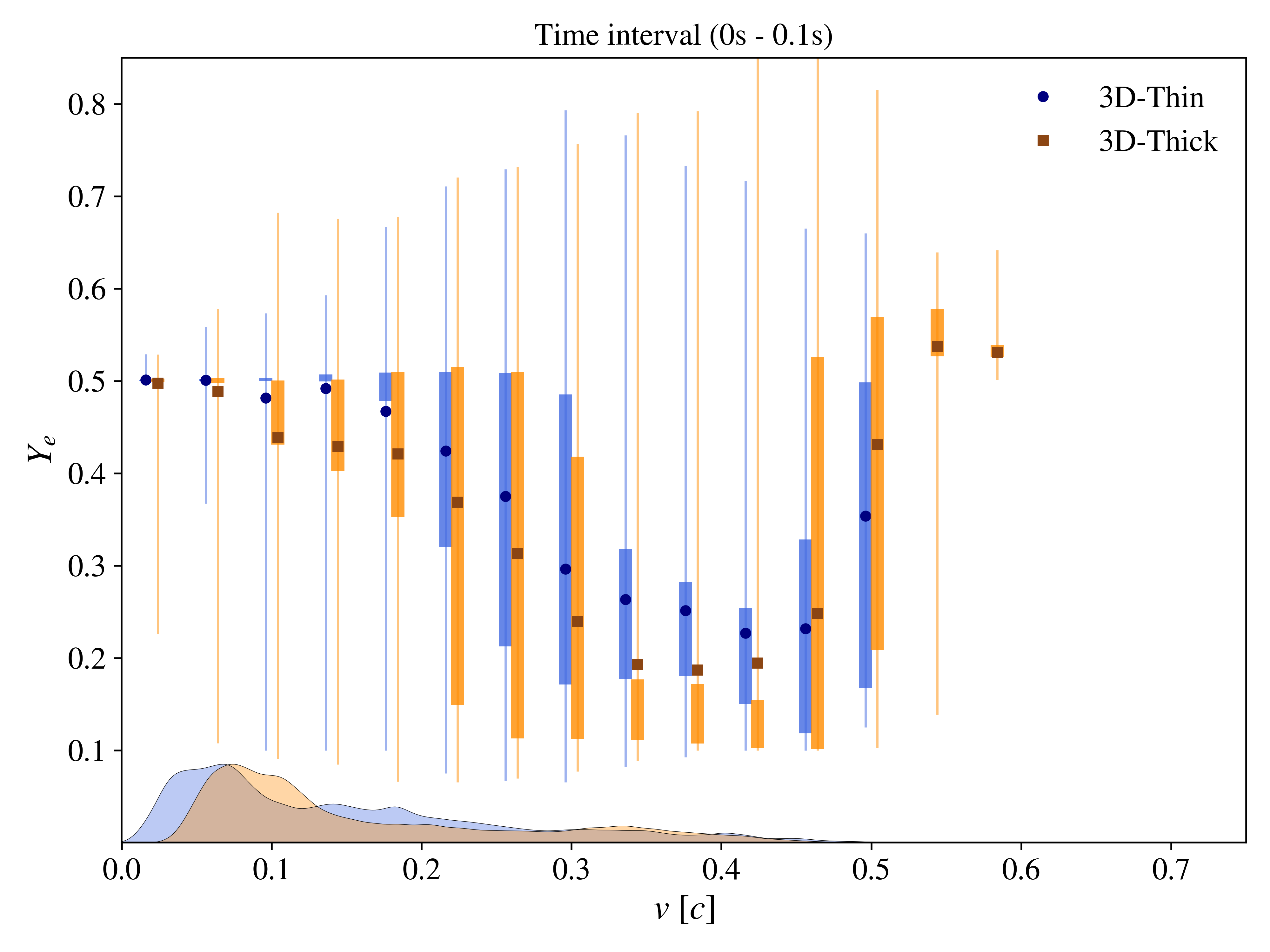}
\includegraphics[width=0.24\textwidth]{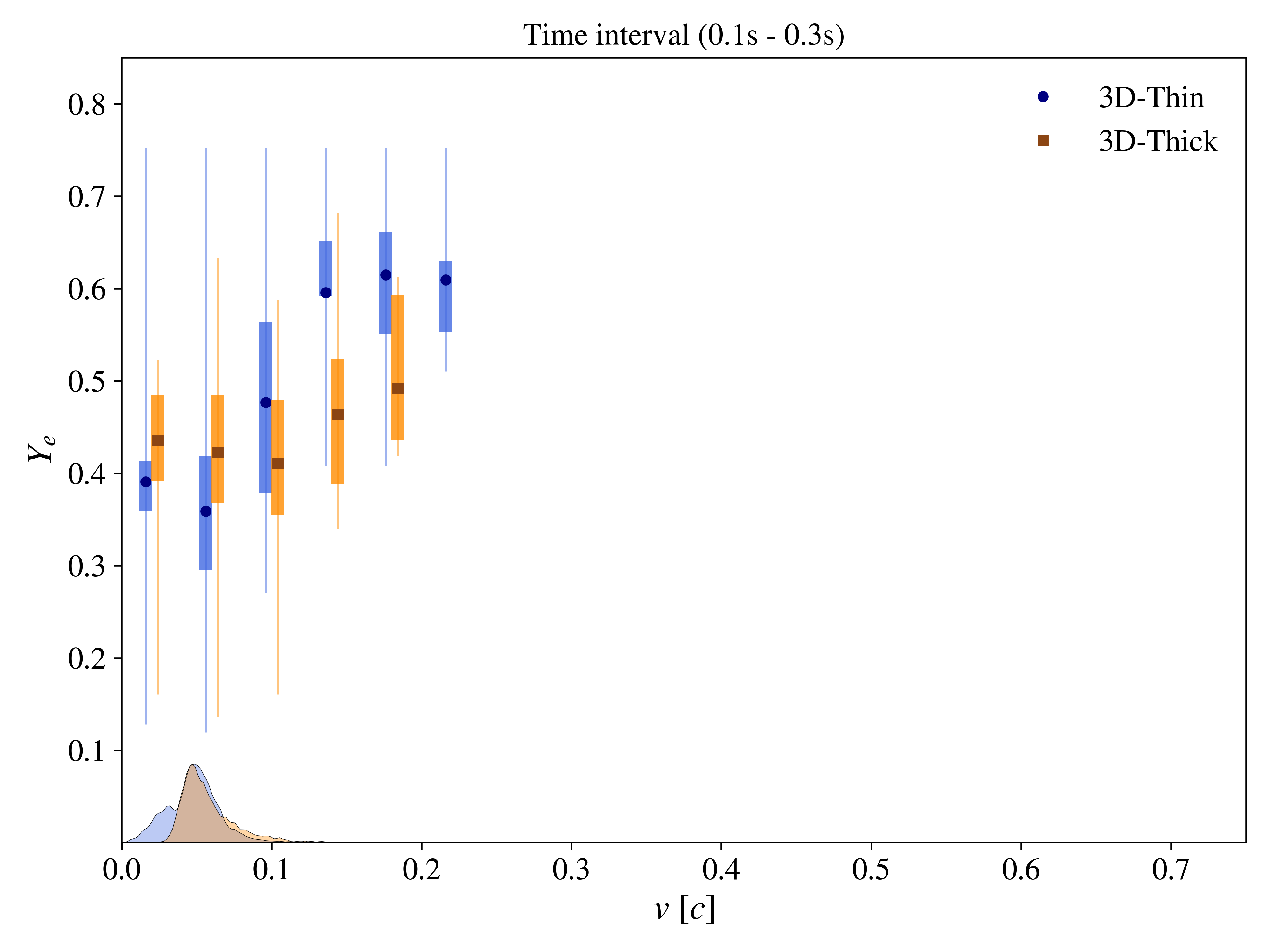}

    \caption{Binned electron fraction ($Y_e$) versus ejecta velocity ($v$), plotted for the 3D models shown in Fig. \ref{fig:Tracers}.
Vertical bars indicate the full range (thin lines) and the interquartile range (thick bars) of $Y_e$ values in each velocity bin, with markers denoting the mean.
The bottom ribbons show the kernel density estimate (KDE) of the velocity distribution for each model.
The left and right panels show ejecta sampled at early and late periods of evolution, respectively.}
    \label{fig:Ye_hist}
  \end{figure}

Figure \ref{fig:Ye_hist} shows the distribution of the electron fraction, $Y_e$, as a function of velocity, v, for all four 3D models, grouped by their BH mass. For each model, we binned the full velocity distribution and plotted the corresponding distribution of $Y_e$ within each velocity bin. The minimum-maximum range of $Y_e$ is indicated with thin lines, the 25\%-75\% quartiles with solid boxes, and the mean values are also marked. To illustrate the temporal evolution, we divided the simulations into early and late phases. For the 3D-Thin and 3D-Thick models (final physical time 0.3 s), these intervals correspond to $0-0.1$ s and $0.1-0.3$ s. For the 3D-kn and 3D-kn-p models (final time 0.8 s), the intervals are $0-0.3$ s and $0.3-0.8$ s. While the velocity is always calculated from tracers, the source of $Y_e$ differs: for the early time analysis, $Y_e$ is evolved directly in the HARM-EOS code, whereas for the late-time analysis, it is taken from the postprocessing results of the nuclear reaction network.

The temporal evolution of the disk wind reveals significant differences in the outflow properties across the models. During the early phase, the ejecta are fast and compositionally inhomogeneous, characterized by a broad $Y_e$ distribution and velocities extending beyond $v = 0.5$c. Within the velocity range with sufficient statistics (see the KDE histogram), we infer that the fastest ejecta are the most neutron-rich, representing fluid elements launched from the internal region of the accretion disk. Although the 3D-Thin, 3D-Thick, and 3D-kn models show similar evolutionary trends, the 3D-kn-p model exhibits a more uniform  $Y_e$ distribution across all velocities. The velocity KDE for 3D-kn-p shows the largest spread in the $0-0.4$c range, suggesting its ejecta are the most dynamically variable. As the wind evolves into the later phase, two trends are observed: the composition settles towards higher $Y_e$ values and the fluid decelerates. The wide $Y_e$ distribution from the early phase tightens considerably, with the mean $Y_e$ reaching $\approx 0.4-0.6$ for the 3D-Thin and 3D-Thick models and $\approx 0.3-0.45$ for the 3D-kn and 3D-kn-p models. The high-velocity component of the 3D-Thin and 3D-Thick wind effectively vanishes, with most late-time material decelerating below $v=0.2$c. By contrast, the 3D-kn-p model maintains a more powerful outflow, with a persistent velocity tail that reaches $v \approx 0.45$c. These results indicate that the central engine of the 3D-kn and kn-p models drives a more sustained and neutron-rich wind compared to the more rapidly waning outflow of the 3D-Thin and 3D-Thick models. They also show that the perturbed model maintains fast outflows even in late times.

\subsection{Abundance patterns and ratios of chosen elements}

We postprocessed the tracers from all nine models and generated isotope abundance profiles using nuclear reaction networks for each model.

\begin{figure}
    \includegraphics[width=0.45\textwidth]{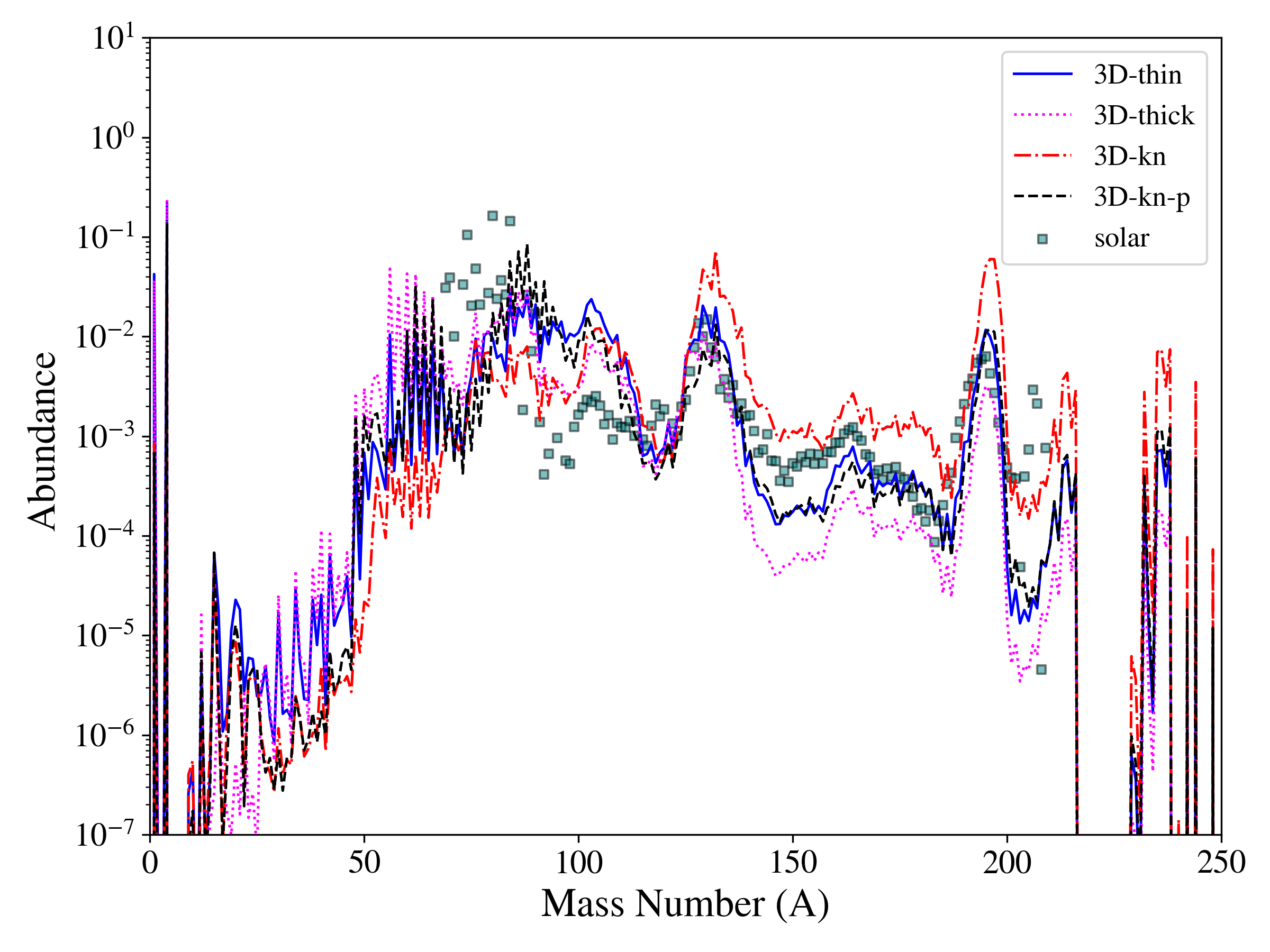}
    \caption{Mass abundance profiles from the nuclear reaction network for all 3D models in the sample. The solar abundance profile is included for comparison.}
    \label{fig:Mass-Abundance}
\end{figure}

Figure \ref{fig:Mass-Abundance} shows the mean relative abundance of isotopes, extrapolated to1~Myr for 3D models, based on all corresponding trajectories. The solar abundance profile \citep{ARNOULD200797} is included in the same plot for comparison.

We performed r-process calculations for all models, postprocessing the tracer particles using SkyNet. To capture the full freeze-out and radioactive decay timescales, we continued the network evolution up to 1 Myr, significantly extending the duration of the simulations. The calculation started once the tracer temperature fell below 10 GK. Initially, we evolved the composition in nuclear statistical equilibrium (NSE); when the temperature dropped to 7GK, the network switched to the standard reaction evolution.

We note that different studies \citep{PhysRevLett.119.231102, Combi_2023} adopt varying thresholds for the onset of network evolution and the NSE transition, typically in the range 5--10~GK. The choice of these transition points influences the results, since the electron fraction $Y_{e}$ at 5~GK is generally higher than at 10~GK, and this difference is reflected in the final abundance patterns  from the full reaction network. For the few tracers whose temperature never exceeded 10~GK, we used the initial values of the required parameters, which were then evolved in the reaction network. The density profile for the late-time evolution was extrapolated following a power-law $\rho\propto t^{-3}$ (homologous expansion). Hence, the abundance profiles reported (Fig. \ref{fig:Mass-Abundance}) represent the asymptotic nucleosynthetic yields, accounting for the cooling and expansion phases in the late period.

The abundance patterns clearly demonstrate robust production of heavy elements through r-process nucleosynthesis, exhibiting the first, second, and third r-process peaks, along with distinct signatures of lanthanides and actinides.  
In the 3D-Thin and 3D-Thick models, the abundances of heavy elements corresponding to the second and third peaks are considerably lower, whereas in the 3D-kn and 3D-kn-p models, this is not the case.
We also note that in all models, light-element abundances increase when the network evolution is initiated at 7 GK.

Although each model exhibits minor variations in the relative abundances of light and heavy nuclei, the overall profiles closely reproduce the Solar abundance distribution and its characteristic r-process features.
We further analyzed these nucleosynthesis patterns by comparing the heavy-to-light element ratios. 
 Figure \ref{fig:Mass-Abundance} highlights clear differences in the nucleosynthetic yields of light and heavy r-process elements across the 3D models. The 3D-Thick model (magenta curve) predominantly synthesizes light r-process nuclei (A < 100), while its production of heavier nuclei beyond the second r-process peak is significantly suppressed. In contrast, the 3D-kn model (dash-dot red curve) generates robust heavy-element abundances, with a pronounced third peak around A $\approx$ 195 and substantial actinide yields, while producing comparatively fewer light r-process nuclei. Specifically, the third r-process peak abundance in model 3D-kn exceeds that of 3D-Thick by approximately an order of magnitude. Conversely, 3D-Thick overproduces light-element nuclei compared to 3D-kn. The remaining two models, 3D-Thin and 3D-kn-p, exhibit intermediate behavior, with abundance patterns lying between these two extremes. Notably, all models reproduce the characteristic r-process peaks near A $\sim$ 80, 130, and 195, although they differ quantitatively in their relative peak amplitudes.

Table \ref{tab:outflows} summarizes the thermodynamic properties of the outflows in various models, along with key characteristics of the resulting nucleosynthesis patterns. The outflows from 2D and 3D models exhibit minor differences in their average electron fraction and nucleosynthetic properties, 
with the most notable discrepancy
observed in the neutrino-thick models. During the expansion phase, as the temperature drops from 10 GK to 5 GK, the average electron fraction in the fiducial models shows minimal variation. 
We observe a more pronounced change in the ``kilonova'' scenario models. Specifically, the \textit{3D-kn-p} model with a non-axisymmetric perturbation develops an electron fraction about twice that of its unperturbed counterpart.

\begin{figure}
    \includegraphics[width=0.45\textwidth]{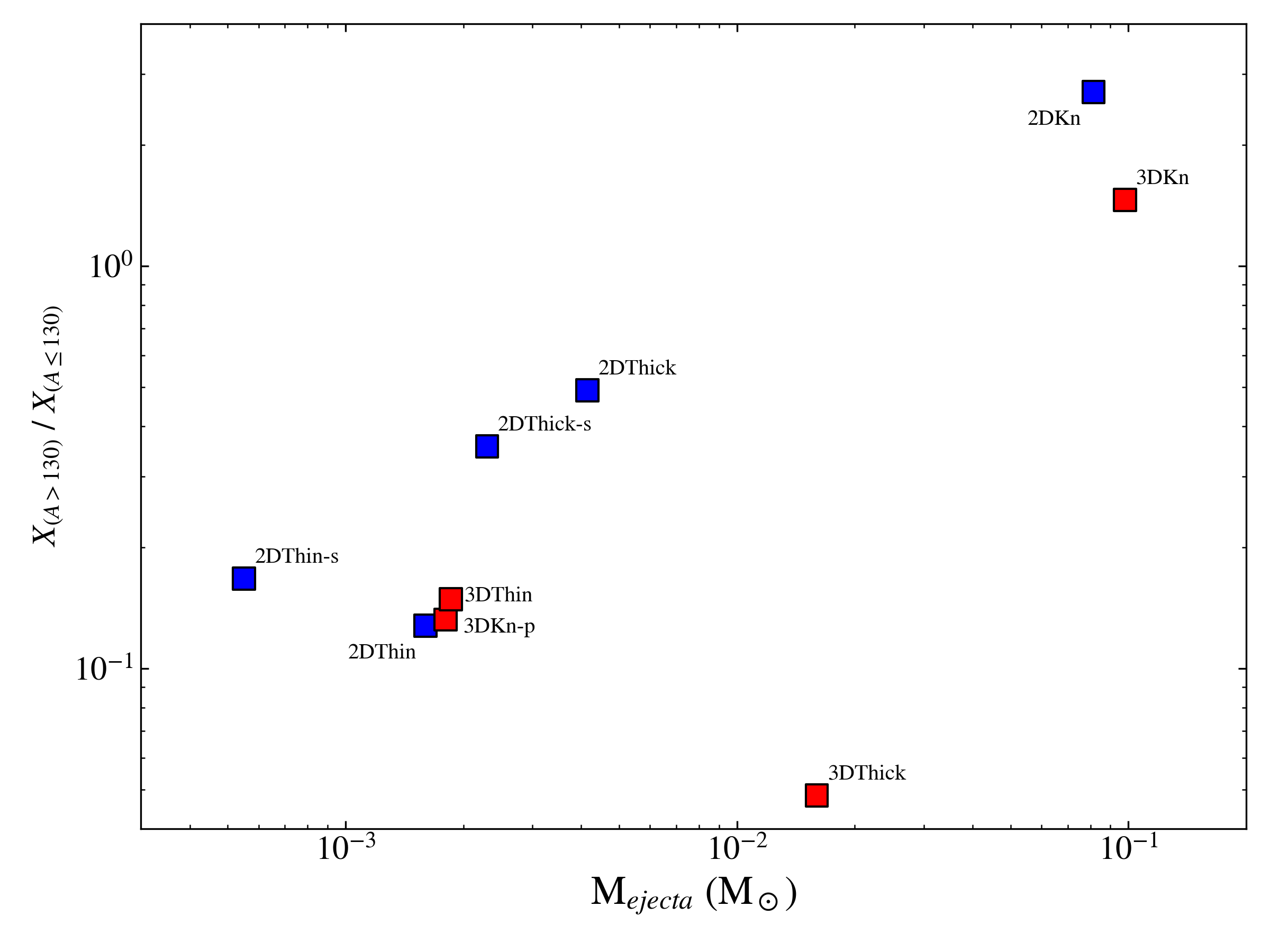}
    \caption{Mass abundance ratios of heavy to light elements from the nuclear reaction network for all models, plotted as a function of the total outflow mass.
    }
    \label{fig:heavy_light}
\end{figure}

Finally, Figure \ref{fig:heavy_light} shows the ratios of heavy-to-light element abundances from our r-process nucleosynthetic calculations. The assumed mass number threshold between ``light'' and ``heavy'' is $A=130$; below this value, the pattern may be highly contaminated by s-process nucleosynthesis, which is beyond the scope of this work. These results are plotted as a function of the total outflow mass, expressed in solar mass units.
The most massive outflows are also the most enriched in heavy elements, with the kilonova models being particularly prominent. 
In contrast, the fiducial thin and thick models produced far fewer ``heavy'' elements relative to ``light'' ones and they have low total ejecta masses. The thick models show a somewhat larger mass ratio of heavy to light elements, but even for the 2D-Thick simulation, the ratio remains below unity.
We also note that the blue points, which represent 2D models, are systematically above the red points, which represent 3D models, suggesting that the dimensionality of the GRMHD simulation has a systematic impact on the abundance profiles.
The correlation between ejecta mass and heavy-element relative abundance is generally visible in both the 2D and 3D simulation sets.

 The only obvious outlier point seen in Fig. \ref{fig:Mass-Abundance} is the 3D-Thick model. In this simulation, the heavy-element ratio in the wind is much smaller than in its 2D representation, at a roughly similar total outflow mass. The 3D-Thick model results in a much larger average electron fraction than the other models ($<Y_{e}>\sim 0.432$, with a shift in $Y_{e}$ from about 0.3 to 0.4, between the 2D and 3D cases. However, in this simulation, the outflow velocity of the bins with large $Y_{e}$ is higher than that in the 2D case. Consequently, the nucleosynthesis computed for high-velocity tracers results in a negligible contribution to the heaviest isotopes. The late-arriving slow tracers have smaller $Y_{e}$, but also carry less mass. We attribute this effect to the magnetically driven wind, which launches outflows from the disk surface (large $Y_{e}$) and from the inner layers (small $Y_{e}$). In our thick models, the magnetic field is initially weaker ($\beta=100$), which may lead to substantial differences in the launching of outflows from various disk layers. 

\begin{table*}[t]
\caption{Outflow thermodynamic properties and r-process nucleosynthesis characteristics.}
\label{tab:outflows}
\centering
\setlength{\tabcolsep}{15pt}           
\renewcommand{\arraystretch}{1.5}    
\begin{tabular}{l c c c c c c}
\hline
Model &
\shortstack{Total outflow\\ mass [$M_\odot$]} &
\shortstack{$\langle Y_e \rangle_{tr}$\\  $T=10\,\mathrm{GK}$} &
\shortstack{$\langle Y_e \rangle_{tr}$\\  $T=5\,\mathrm{GK}$} &
\shortstack{$\langle v \rangle$[c] \\ at $500\,r_{\rm g}$} &
$Eu/Fe$ &
\shortstack{$X(57\le Z \le 72)$\\ (Lanthanides)} \\
\hline\hline
2D-Thick-s & $2.3 \times 10^{-3}$ & 0.388 & 0.424 & 0.125 & $8.2 \times 10^{-3}$ & 0.0011 \\
2D-Thin-s  & $5.5 \times 10^{-4}$ & 0.382 & 0.444 & 0.089 & $4.6 \times 10^{-3}$ & 0.0007 \\
2D-Thick   & $4.2 \times 10^{-3}$ & 0.328 & 0.389 & 0.097 & $1.7 \times 10^{-2}$ & 0.0020 \\
2D-Thin    & $1.6 \times 10^{-3}$ & 0.378 & 0.448 & 0.111 & $5.5 \times 10^{-3}$ & 0.0007 \\
2D-kn      & $8.2 \times 10^{-2}$ & 0.160 & 0.189 & 0.073 & 1.04               & 0.0509 \\
\hline
3D-Thick   & $1.6 \times 10^{-2}$ & 0.420 & 0.485  & 0.132 & $8.3 \times 10^{-4}$ & 0.0003 \\
3D-Thin    & $1.9 \times 10^{-3}$ & 0.377 & 0.445  & 0.100 & $1.2 \times 10^{-2}$ & 0.0008 \\
3D-kn      & $9.8 \times 10^{-2}$ & 0.191 & 0.262  & 0.074 & $5.7 \times 10^{-1}$ & 0.0265 \\
3D-kn-p    & $1.8 \times 10^{-3}$ & 0.372 & 0.420  & 0.193 & $4.6 \times 10^{-2}$ & 0.0016 \\
\hline
\end{tabular}
\tablefoot{Model parameters are listed in Table~1. The total mass and angle-averaged electron fraction are computed for unbound tracers cooled to the NSE-breakdown temperature. The velocity (in units of $c$) is measured at $500\,r_{\rm g}$. Europium-to-iron number ratios are calculated across all isotopes (not normalized to Solar). The lanthanide-abundance fraction is summed over $57\le Z\le72$. 
}
\end{table*}

\section{Discussion}
\label{sect:discussion}

\subsection{Heavy-element production across binary merger scenarios}

Compact binary mergers are the main sites of r-process heavy-element production. However, the detailed outcome of the process is highly dependent on the merger progenitor specific type. Binary neutron star (NS-NS) mergers produce more massive ejecta, with the neutron-rich component typically having a mass of $\sim 0.01-0.05 M_{\odot}$ \citep{Matsuno2021}. (More massive ejecta generally occur in BNS under favorable conditions; however, BH-NS mergers can sometimes produce even larger ejecta\cite{kruger_foucart2020})
In contrast, BH-NS merger ejecta may be less massive, or even negligible, with outflows of $\sim 0-0.02 M_{\odot}$ as reported by various authors \citep{Drozda2022, Kobayashi2023}. 
In our models, the post-merger systems produce total masses of unbound outflows in the 2D-Thin-s, 2D-Thin, and 3D-Thin models that are systematically lower than in the other scenarios; these represent the BH-NS post-merger configurations.
Electromagnetic counterparts are rarely expected from these events, with only $\sim 1 \%$ of these mergers expected to produce detectable kilonovae \citep{Drozda2022}.

Signals are expected mainly from NS-NS merger events. According to LIGO/Virgo estimates, their frequency is $\sim 10-1700$~Gpc$^{-3}$yr$^{-1}$ \citep{Abbott2023}, corresponding to roughly one event per $10^{4}$ years per galaxy. Comparison with the BH-NS merger rate shows that the number density is 1-2 orders of magnitude lower.

The overall contribution of binary mergers to the chemical enrichment of galaxies and stars is, however, affected by their typical delay times. As reported by \cite{Kobayashi2023}, BH-NS mergers may have shorter delay times because their more massive progenitor stars evolve rapidly to form the BH companion. The NS-NS binary mergers occur over a broad range of delay times extending up to gigayear scales \citep{skulado2020}.
 However, growing evidence indicates a distinct fast-merging population with merger times as short as 10–100 Myr, which could enrich metal-poor environments at early epochs 
\citep{beniamini2019, Maoz_2025}.
While these fast mergers contribute crucially to early r-process enrichment, long-delay mergers also dominate the overall galactic budget. Therefore, NS-NS mergers alone cannot fully explain the r-process abundances observed in the very first stars. 
Nevertheless, NS-NS mergers appear to be the dominant producers of heavy r-process elements in the Milky Way\citep{Chen2024}, whereas BH-NS mergers are significant only in certain niche cases with favorable conditions for launching more massive ejecta.

\subsection{Chemical enrichment due to mergers}
\label{sect:archeo}

Galactic archaeology studies the structure and evolution of the Milky Way and other galaxies by analyzing stellar populations, chemical abundances, and stellar kinematics, which serve as fossil evidence of their history \citep[e.g.,][]{Farouqi2025}. Although galactic chemical enrichment involves multiple sources such as supernovae and winds from massive stars \citep{2023MNRAS.522.3092L}, compact binary mergers are considered a primary site for the production of heavy r-process elements ($A \ge 120$; \cite{Kobayashi2023}). Traces of this ancient enrichment are found in the atmospheres of extremely metal-poor stars, whose compositions reflect the yields of individual nucleosynthesis events from the early Universe \citep{hansen2020}.

Ejecta from a BNS or BH-NS merger are complex, consisting of both early-time dynamical ejecta and later-time winds from a post-merger accretion disk. The simulations presented here focus on this disk wind component. Our suite of models shows that these winds can eject between $M \sim 0.001- 0.1 M_{\odot}$ (see Table \ref{tab:outflows}) of r-process material per event. The final nucleosynthetic yields are sensitive to the properties of the central engine and disk. For instance, our models with a lower disk-to-BH mass ratio (e.g., 3D-Thin) may represent BH-NS merger scenarios and produce a specific abundance pattern. The ejecta are also sufficiently neutron-rich to synthesize a full complement of heavy elements, including a significant lanthanide abundance fraction ($X_{Lan} \sim 10^{-2}-10^{-4}$).

A relevant question is how the yields from our models can account for the Galaxy's r-process enrichment. Considering a Milky Way BNS merger rate of $\sim$ 20-40 Myr$^{-1}$ \citep{Colom_2023,10.1093/mnras/stad2768}, the ejecta masses of $M \sim 0.001- 0.1 M_{\odot}$ per event found in our simulations are broadly sufficient to explain a dominant fraction of the Galactic r-process material. A robust assessment, however, requires detailed Galactic chemical evolution (GCE) models \citep[e.g.,][]{Chen_2025}, which can self-consistently incorporate cosmological event rates, merger delay-time distributions, and inhomogeneous mixing of ejecta into the star-forming interstellar medium.

The relative contributions of various progenitor channels, including alternatives such as NS-WD mergers, are also being actively considered \citep{Kaltenborn_2023, Liu_2025, Chen2024}, although their nucleosynthetic yields remain uncertain. Within this context, our work provides physically-grounded nucleosynthetic yields for the post-merger disk wind channel. 
Even within this single channel, we find substantial intrinsic variability in key nucleosynthetic outflow properties; ejecta mass, $\langle Y_e \rangle$, and velocity, which yield a wide range of abundance ratios, such as Eu/Fe and lanthanide abundance fractions (see Table \ref{tab:outflows}). This intrinsic yield diversity may provide a natural physical origin for at least part of the chemical-abundance scatter observed among metal-poor halo stars \citep{Cescutti2015,Wehmeyer2015}. Our results suggest that the observed diversity of r-process enrichment does not necessarily require multiple distinct astrophysical sites; it may instead reflect the intrinsic diversity of merger and disk configurations coupled with the stochastic mixing of their ejecta into the early Galactic medium.
These yields provide a critical and often uncertain input for GCE frameworks, enabling tests of whether the observed diversity of r-process patterns is a natural outcome of the variety of merger events.

\subsection{Source 211211A}
\label{sect:211211}

The unprecedented discovery of a long-duration (50~s) GRB 211211A associated with a kilonova indicates that this event may be associated with a BNS merger. The extended emission of the gamma-ray burst exhibited a hard GRB, followed by a prolonged, softer emission \citep{Rastinejad2022}. The kilonova associated with this event was similar in terms of luminosity, duration, and color to that observed in the prototypical event GW-GRB 170817, whose progenitor was confirmed to be a BNS merger.
However, the 211211A source also exhibited several LGRB characteristics, including a hardness ratio near their mean value, and no associated supernova signal was detected. Thus, a massive-star origin has been ruled out for this source.

Multiwavelength observations indicate an isotropic-equivalent energy of  $5\times 10^{52}$ erg \citep{2023NatAs...7...67G}, placing it at the higher end of the GRB luminosity distribution \citep{Kumar2015}.
The afterglow emission was fitted using an additional three-component kilonova model to explain the excess K-band brightness and optical-NIR data. From the fits, we determined the total r-process ejecta mass to be 0.047 $M_{\odot}$. The red, purple, and blue ejecta were fitted with velocities of 0.3, 0.1, and 0.3\textbackslash{},c, respectively.

Our scenario for this source assumes an engine formed after the BNS merger, consisting of a rotating BH with a mass of 8.2 $M_{\odot}$ and spin $a=0.6$, surrounded by a remnant accretion disk that is ten times less massive than a BH. This engine produces a powerful, magnetically dominated jet responsible for the GRB prompt emission and also generates a substantial mass outflow, which may be the source of the ``red'' kilonova component. The source of opacity is due to a significant mass fraction of lanthanides in this scenario, which is 100 times greater than in our other fiducial simulations.
This scenario is therefore consistent with long binary GRBs recently discussed by \cite{Gottlieb2025}, which originate from the accretion of massive ($M\ge 0.1 M_{\odot}$) disks onto a BH formed after the transient hypermassive neutron star collapses to a BH. 
 The extended GRB duration in this case results from prolonged accretion prior to reaching the magnetically arrested state. However, this state does not form in our simulations, 
while the lanthanide abundance fraction is significantly higher in 2D-kn and 3D-kn than in the other fiducial models. We speculate that the 
longer GRB prompt phase might originate from a potentially eccentric BH-NS merger, as tested by the perturbed model 3D-kn-p (see Sect. \ref{sect:duration}.

\subsection{A long-duration post-merger GRB}
\label{sect:duration}

The duration of a GRB ($t_{\rm GRB} \sim t_{\rm CE}$) can be determined by how long the central engine powers the jet \citep[e.g.,][]{Beniamini_2020,Zhang2025}. The jet power
$P_{\rm j}= \eta\dot{M}c^2 \sim t$ remains roughly constant until the model reaches the MAD state; the extended GRB duration results from prolonged accretion prior to the MAD state.
Central engine activity decreases after the MAD state; when $P_j \sim t^{-2}$, then $t_{\rm GRB} \sim t_{\rm MAD}$. 

A long-duration GRB ($t_{\rm GRB} \sim 10$~s) can result from a post-merger thick disk with mass $M_d = 0.1\,M_\odot$ according to \citep{Gottlieb2023a}. 
Using semi-analytic models, \citet{LuQuataert2023} show that thinner disks with masses in the range $10^{-3}$-$10^{-2}\,M_\odot$ can also produce extended-emission GRBs due to the slow evaporation of the accretion disks.

Our perturbed model, 3D-kn-p, may explain an extended-emission GRB.
The model does not exceed the MAD threshold (Section~\ref{sec:fiducialThinAndThick}). In addition, the red kilonova ejecta reported in Table~\ref{tab:outflows} falls within the same range as that presented  
by \citet{Rastinejad2025} for GRB~230307A, 
which shows a red ejecta mass on the order of $10^{-3}\,M_\odot$, with low escape velocities ($v \sim 0.1c$).
Although 
these authors report different relationships between GRB duration and ejecta mass, they
suggest that LGRBs arise from
mergers. Quantitatively, our perturbed model 
satisfies the key characteristics discussed above.
    
\subsection{The influence of magnetized winds on neutrino emission and the orphan kilonova} 

Our results indicate that the neutrino luminosity is high in models with a low magnetic flux and/or low initial magnetization. This dominance of neutrino emission persists even in moderately magnetized disks.
The weakly magnetized 3D-Thick model, initialized with $\beta=100$, shows an early peak in magnetic flux that later decays as $\Phi_{\rm BH} \propto t^{-1}$. At $t = 300$~ms, the magnetic flux decreases to $\Phi_{\rm BH}\sim10^{27}$~G~cm$^{2}$, while the neutrino luminosity reaches $L_{\nu}\sim10^{51}$~erg~s$^{-1}$.
The more magnetized model, 3D-Thin, starting from $\beta=50$, maintains an almost constant magnetic flux, $\Phi_{\rm BH}\propto t^{-0.1}$ and exhibits neutrino luminosity $L_\nu \sim 2\times10^{50}$~erg~s$^{-1}$. This model is more magnetized but about an order of magnitude less massive than the 3D-Thick case. The stronger magnetization drives powerful winds, reducing the volume of the neutrino-emitting region and decreasing its thermal energy. 
The higher magnetized model 3D-kn, also initialized with $\beta=50$, presents nearly constant magnetic flux of $\Phi_{\rm BH}\sim10^{28}$~G~cm$^{2}$ ($\propto t^{-0.1}$) and luminosity $L_{\nu}\sim2.3\times10^{51}$~erg~s$^{-1}$. Despite its relatively low spin ($a=0.6$), this model produces strong mass outflow and intense neutrino emission.
In contrast, the perturbed model 3D-kn-p shows that magnetization does not strongly influence neutrino output. The magnetic flux remains constant but lower ($\Phi_{\rm BH}\sim10^{22}$~G~cm$^{2}$), and the neutrino luminosity is $L_{\nu}\sim10^{51}$~erg~s$^{-1}$. The efficiency of neutrino annihilation can in this case be larger than about 1\% \citep{2019MNRAS.482.2973D}; however, we speculate that this scenario leads to a weak GRB or an orphan kilonova without a jet.

None of our 3D models reach the magnetically arrested disk (MAD) state; therefore, strong Poynting-dominated jets are absent, 
despite high-magnetization funnels ($\sigma \sim 50$) forming at later stages.
The neutrino emissivity is mainly dependent on the internal energy of the disk. As long as winds are not strongly enhanced by magnetization, neutrino luminosity remains dominant. When the mass ejection increases, it proceeds smoothly; thus, the thermal energy does not vary abruptly. Based on the estimated 
mass loss, we find that magnetized disks can produce both strong winds and significant neutrino emission. 

In several cases, the magnetization decreases with time, while the neutrino luminosity becomes dominant.
Even when the jet becomes weakly magnetized, it can still successfully penetrate the post-merger environment. 
Jet efficiency does not appear to be strongly affected by the degree of magnetization in our initial conditions, implying that jet launching occurs even in low-magnetization scenarios.  
\citet{Ruiz2017} and \citet{Hayashi2025} show that, after a binary merger, the neutrino energy flux can exceed the magnetic energy flux by up to three orders of magnitude. This dominance challenges jet-launching mechanisms, especially those that rely on Poynting-driven jets \citep[e.g.,][]{Aguilera-Miret2024}. In contrast, collapsar models show the opposite behavior: strong magnetic flux initially is required to penetrate the stellar envelope \citep{Burrows_2007,KomissarovBarkov2009} and magnetic flux decreases over time while the neutrino luminosity maintains ram pressure once the jet has cleared the dense material \citep[e.g.,][]{ObergaulingerAloy2020}.

\section{Conclusions}
\label{sec:conclusions}

This work explores the properties of the GRB central engine in post-merger scenarios, focusing on neutrino-cooled, magnetically driven winds from an accretion disk around a BH.
The system formed after the binary compact merger event may contribute significantly to r-process nucleosynthesis and observable kilonova emission, while also chemically enriching the host galaxy.
 The main outcomes of our modeling are as follows:
 \begin{itemize}
     \item We obtain plausible scenarios for short GRB engines with prompt jets powered by neutrino annihilation. 
     \item We confirm that the scenario can model weak jets and magnetically driven wind outflows.
     \item To model a sufficient mass loss in the outflows and thereby significant wind contributions to the kilonova signal, massive disks are required.
    \item The kilonova can be modeled using moderate BH spin; however, the SANE accretion mode in this model produces low power for the Poynting-dominated jets and is therefore unable to reproduce the bright source GRB211211A.
    \item Our simulation reveals significant intrinsic diversity in nucleosynthesis yields across different disk configurations, with variations in $Y_e$, ejecta mass, and velocities, leading to distinct abundance patterns of light and heavy elements.
   \item We quantify the Eu/Fe abundance ratio as a key tracer of Galactic chemical enrichment and note that kilonovae powered by disk winds represent an important channel contributing to this process.

 \end{itemize}

\begin{acknowledgements}
We thank Ariadna Murguia-Berthier, Oleg Korobkin and Nicole Lloyd-Ronning for helpful discussions. We also thank Piotr Plonka for help in calculating BZ luminosities.
  The project was partially supported by grant 2019/35/B/ST9/04000 from Polish National Science Center.AJ was additionally supported by grant 2023/50/A/ST9/00527.
  We also acknowledge Cyfronet AGH via grant  PLG/2024/017013.
  We acknowledge Polish high-performance computing infrastructure PLGrid for awarding this project access to the LUMI supercomputer, owned by the EuroHPC Joint Undertaking, hosted by CSC (Finland) and the LUMI consortium through PLL/2024/07/017501
   The research was conducted with the support of the Interdisciplinary
Center for Mathematical and Computational Modeling of the University of Warsaw (ICM UW) under computational allocation no g100-2230.
\end{acknowledgements}

\bibliographystyle{aa}
\bibliography{aanda}

\begin{appendix}
\section{Recovery method}
\label{sect:recovery}

We use the EOS adapted from Helmholtz tables, with $P(\rho, T, Y_{e})$ and $\epsilon(\rho, T, Y_{e})$, where $Y_{e}$ is the electron fraction. It is usable for a wide range of densities and temperatures. The evolving electron fraction is giving an additional source term to the energy equation.

The GRMHD code solves system of equations in the form
\begin{equation}
\del_t \bU(\bP) = -\del_i \bF^i(\bP) + \mathbf{S}(\bP)
\label{eq:scheme}
\end{equation}
where $\bU(\bP)$ are conserved variables that
evolve through the fluxes at cell boundaries and source terms.
Here $\bP$ 
are the 'primitive variables, which are subject to the EOS.

Explicit form of primitive and conserved
variables, fluxes, and source terms, is:

\begin{equation}
\bP = [\rho, ~ \sB^{k}, ~ \tu^{i}, ~ Y_{e}, ~ T]   \nonumber
\end{equation}
\begin{equation}
\bU (\bP) = \sqrt{-g}[\rho u^{t}, ~ T^{t}_{t}+\rho u^{t}, ~ T^{t}_{j}, ~ B^{k}, ~ \rho Y_{e}u^{t}]  \nonumber
\end{equation}
\begin{equation}
\bF^{i}(\bP) = \sqrt{-g}[ \rho u^{i}, ~ T^{i}_{t}+\rho u^{i}, ~ T^{i}_{j}, ~ (b^{i} u^{k} - b^{k}u^{i}), ~ \rho Y_{e} u^{i} ] \nonumber
\end{equation}
\begin{equation}
\bS (\bP) = \sqrt{-g}[0, ~ T^{\kappa}_{\lambda}\Gamma^{\lambda}_{t \kappa} + \sQ u_{t}, ~  T^{\kappa}_{\lambda}\Gamma^{\lambda}_{t \kappa} + \sQ u_{i}, ~ 0, ~ \sR]
\nonumber
\end{equation}

Here $B^{i}=\sB^{i}/\alpha = \dF^{it}$ is magnetic field three-vector, and $\tu^{\mu}=(\delta^{\mu}_{\nu}+n^{\mu}n_{\nu})u^{\nu}$ is the projected four-velocity, where the orthogonal frame velocity is $n_{\mu}=[-\alpha,0,0,0]$, $n^{\mu}=[1/\alpha,-\beta^{i}/\alpha]$, with lapse 
$\alpha=1/\sqrt{g^{tt}}$, and shift function $\beta^{i}=-g^{ti}/g^{tt}$.

In non-relativistic MHD, both $\bP \rightarrow \bU$ and  $\bU
\rightarrow \bP$ have a closed-form solution. In GRMHD 
$\bU(\bP)$ is a complicated, nonlinear relation. Inversion
$\bP(\bU)$ is calculated 
once or twice in every time step, 
numerically, by the recovery scheme.

There exist a number of inversion schemes, based on specific transformations between these five
independent variables 
\citep{siegel2018}. Inversion is a complex procedure specifically for a non-adiabatic relation of the pressure with density (ie. if the EOS is not given by the so called gamma-law, $p(\rho) = (\gamma-1)\rho\epsilon$).
The numerically computes tabulated
EOS can be of general form of
\begin{itemize}
\item 1-parameter $\epsilon(\rho)$, $P_{\epsilon}(\rho)$
\item 2-parameter $\epsilon(\rho, T)$, $P_{\epsilon}(\rho, T)$
\item 3-parameter $\epsilon(\rho, T, Y_{e})$, $P_{\epsilon}(\rho, T, Y_{e})$
\end{itemize}
In the former work \cite{Janiuk2019}, we used the code based on the 2-parameter tabulated EOS. In the current work, we use the 3-parameter one.

We need to recover the primitive variables, as they are required to construct $T^{\mu\nu}$ and flux terms in fluid evolution.
Recovery methods can lead to bounded or unbounded solution, and the typically used 
Newton-Raphson methods are unbounded.
They exhibit a faster convergence but are less stable, therefore in this work, for some parameter range, we switch to the bracketed root-finding method of \cite{palenzuela2015}
which is slower but more robust. We combined these two methods in our code, to effectively obtain minimum errors for a wide range of temperatures and densities.

\begin{figure}
    \includegraphics[width=0.49\textwidth]{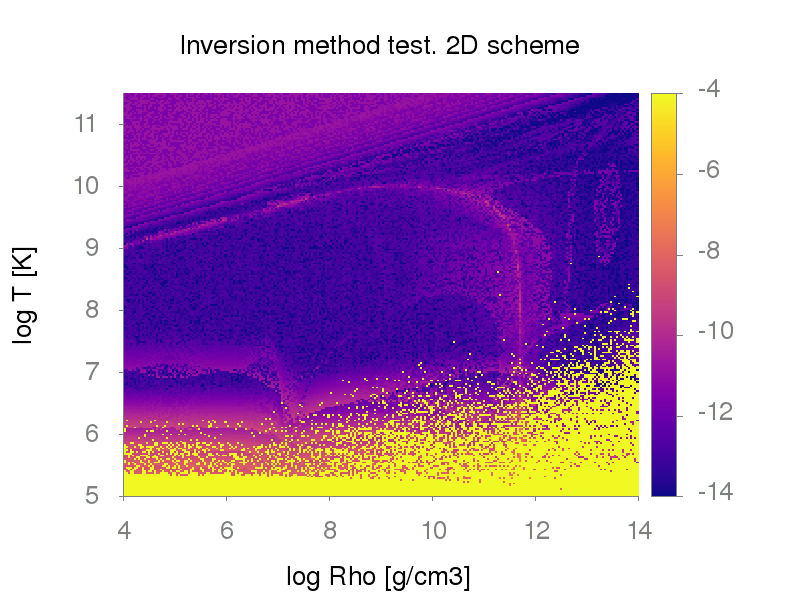}
    \includegraphics[width=0.49\textwidth]{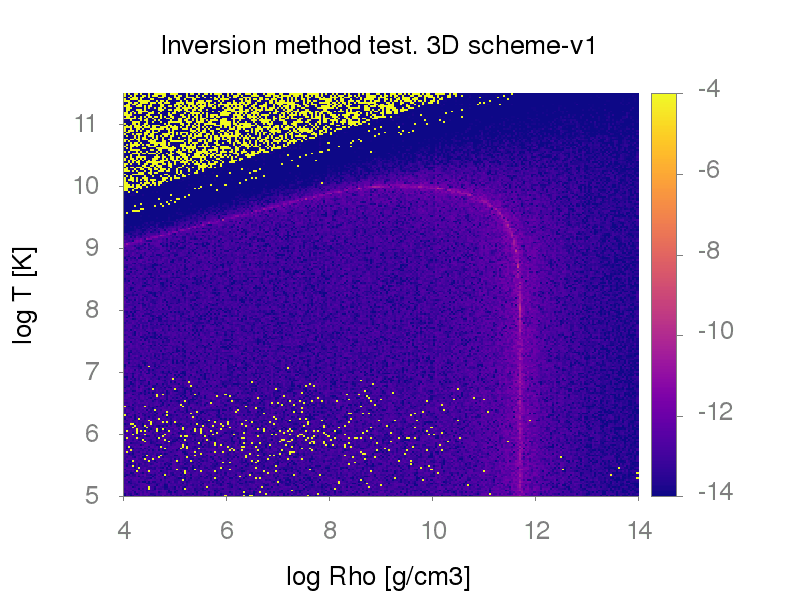}
    \includegraphics[width=0.49\textwidth]{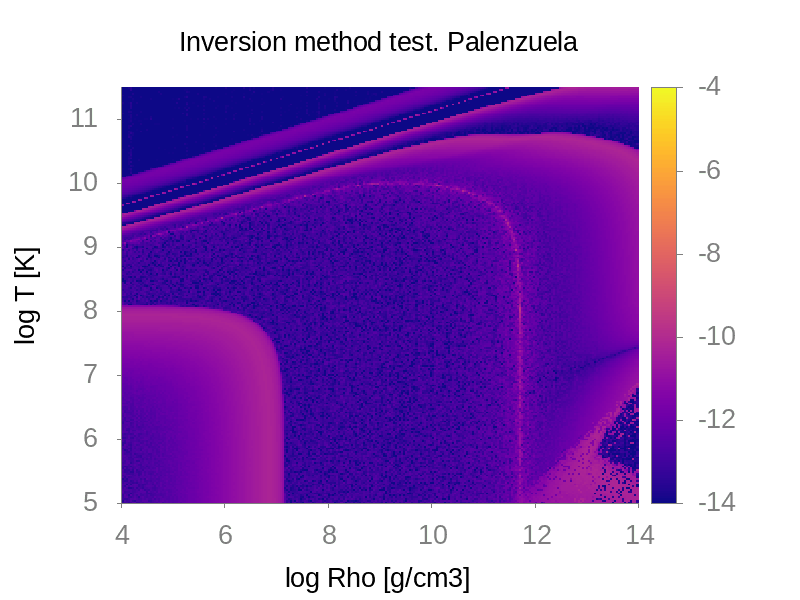}
    \caption{Convergence tests results for the 2D, 3D recovery scheme, and Palenzuela scheme.
    }
    \label{fig:recovery}
  \end{figure}

In Figure \ref{fig:recovery} we present a comparison of our convergence tests made for the 3D scheme and the Palenzuela scheme.
The 3D scheme (method-v1) computes specific internal energy from state vector $x$ and conservatives
  as in Eq. (25) in Cerda-Duran et al. (2008). We solve $f_{3}$, as the system is reduced to solve 3 equations, on $\Gamma, z$, and $T$, by adding a constraint on the internal energy given by EOS tables, namely:
\\
\[ f_{1}: ~~ [\tau + D -z - B^{2} + {B^{i}Q_{i} \over 2 z^{2}} + \rho ] \Gamma^{2} - {B^{2} \over 2} = 0 \]
\[ f_{2}: ~~ [(z+B^{2})^{2} - Q^{2} - {{2z +B^{2}} \over {z^{2}}} (B^{i}Q_{i})^{2}] \Gamma^{2} - (z+B^{2})^{2} = 0 \]
\[ f_{3}: ~~ \epsilon - \epsilon(\rho, T, Y_{e}) = 0 \] 
Here the temperature is employed directly as an unknown through
$ \epsilon(\Gamma, z, T) = h -1 - {P \over \rho} = {z-D\Gamma - \rho \Gamma^{2} \over D\Gamma }$
and does not require inversion of the EOS.
Notice here notation: 
$ Q_{\mu} = -n_{\mu}T^{\nu}_{\mu} = \alpha T^{t}_{\mu} $
is energy-momentum density,
$\Gamma$ is Lorentz factor, $z$ is enthalpy, $ D= -\rho n_{\mu} u^{\mu} = \alpha \rho u^{t}$ is comoving density, and $\tau = -(n_{\mu}n_{\nu} T^{\mu \nu} +D)$.

In 'Palenzuela' scheme, the code is solving 1D equation, for the rescaled variable
\[ \chi = {{\rho h \gamma^{2}}\over {\rho \gamma} }\]
Other quantities are also rescaled, accordingly, to give Lorentz factor, and we give the brackets for $\chi$:
\[ 2 - 2 {{Q_{\mu}n^{\mu} +D}\over {D}} - {\sB^{2}\over D} < \chi <  1  - {{Q_{\mu}n^{\mu} +D}\over {D}} - {\sB^{2}\over D} \]
The equation
\[ f(\chi) = \chi - \tilde\gamma (1+\tilde\epsilon + {\tilde P \over \tilde\rho}) = 0 \]
is solved, with $\tilde{P} = P(\tilde\rho, \tilde\epsilon, Y_{e})$ found in tables.

The tests presented in Figure \ref{fig:recovery} were performed over a broad range of temperatures and densities. We show the value of relative error, summed over primitive
variables, recovered through the alternative schemes.
The parameters used were:  electron fraction $Y_{e}=0.1$, Lorentz factor $\Gamma=2$, gas-to-magnetic pressure ratio $p_{gas}/p_{mag}=10^{5}$.
Conserved variables were derived in Kerr metric, then the primitives were perturbed by a factor of 1.05.
The variables recovered through the recovery scheme are compared to the unperturbed, to calculate error of 
$Err = \Sigma_{k=0, NPR} (P_{k} - \bar{P_{k}})^{2}$. 

For the 3D scheme, we checked two versions,  and the test shown in the plot presents better convergence. As an alternative, instead of the specific internal energy, we can compute pressure from state
vector $x$ and conservatives and then solve for $f_{3}$, as done e.g., in \cite{2021ApJ...919...95M}. 
This version however good for high temperatures, showed bad convergence for temperatures below $T\le 10^{7}$ K. Similarly, the 2D method presented in \cite{2021ApJ...919...95M} showed bad convergence at low temperatures 
however in our case the method is satisfactory for most of temperature range at low densities. 
The bracketed 'Palenzuela' scheme is behaving well, and was used at high temperature and low density regime.

\section{Neutrino treatment}
\label{sect:neutrino}

The neutrino leakage scheme computes a gray optical depth estimate along radial rays for electron neutrinos, antineutrinos, and heavy-lepton neutrinos (nux),
\begin{equation}
\tau(r,\theta,\phi) = \int_{r}^{R} \sqrt{\gamma_{rr}}\bar\kappa_{\nu_{i}}
\end{equation}
and then computes local energy and lepton number loss terms, $\sQ$ and $\sR$, that are evolved via source terms by the GRMHD scheme Eq. \ref{eq:scheme}.

The scheme is based on equations provided by \cite{rosswog2003}, giving the neutrino number emission rates and energy emission rates per unit volume. These effective rates account for locally produced neutrinos as well as the diffusive terms.
The source code of the leakage scheme has been downloaded from the website\footnote{
{\it https://stellarcollapse.org}} and we implemented it in our GRMHD code,
HARM\_EOS, where it is coupled to the inversion procedures. The scheme captures effects of deleptonization, neutrino cooling and heating and enables approximate predictions for the neutrino luminosities. Details and implementation to the core-collapse supernovae simulations can be found in \cite{2012PhRvD..86b4026O}.

\end{appendix}

\end{document}